\documentstyle[aps,eqsecnum]{revtex}
\tightenlines
\begin{document}
%
\twocolumn[\hsize\textwidth\columnwidth\hsize\csname
@twocolumnfalse\endcsname 
\title{Magnetoresistance of a  two-dimensional electron gas with 
  spatially periodic lateral modulations: Exact consequences of Boltzmann's
  equation} 
\author {Rolf Menne and Rolf R.\ Gerhardts}
\address{
Max-Planck-Institut f\"ur Festk\"orperforschung, Heisenbergstrasse 1,
D-70569 Stuttgart, Federal Republic of Germany
}
\date{29 August 1997}
\maketitle
\widetext
\begin{abstract}
\leftskip 54.8pt
\rightskip 54.8pt
On the basis of Boltzmann's equation, and including anisotropic scattering in
the collision operator, we investigate the effect of one-dimensional
superlattices on  two-dimensional electron systems. In addition to
superlattices defined by static electric and magnetic fields, we consider
mobility superlattices describing a spatially modulated density of scattering
centers. We prove that magnetic and electric superlattices in $x$-direction
affect only  the resistivity component $\rho_{xx}$ if the mobility is
homogeneous, whereas a mobility lattice in  $x$-direction in the absence of
electric and magnetic modulations affects only $\rho_{yy}$. Solving
Boltzmann's equation numerically, we calculate the positive magnetoresistance
in weak magnetic fields and the Weiss oscillations in stronger fields within
 a unified approach.
\end{abstract}
\leftskip 54.8pt
\rightskip 54.8pt
]
%
\narrowtext
\section{Introduction}
Two-dimensional electron systems (2DESs) in high-mobility $\rm
Al_{1-x}Ga_x$As-GaAs heterostructures with one-dimensional lateral
superlattices show interesting magnetotransport properties
at low temperatures ($T \approx 4$K or less) 
\cite{Weiss89:179,Gerhardts89:1173,Winkler89:1177}. 
Experiments on typical samples with electron density $n_{el} \sim 3 \cdot
10^{11}$ cm$^{-2}$, mobility $\mu \sim 10^6 $ cm$^2$/Vs,  and  a weak electric
modulation \cite{Weiss89:179} in $x$-direction of period $a \sim 300$ nm
 revealed the following features. If the
homogeneous magnetic field $B_0$ applied perpendicular to the 2DES is very
weak  ($B_0 \leq 0.03$T), one finds a pronounced positive magnetoresistance
 in the component $\rho_{xx}$. With increasing $B_0$, this saturates and is
followed by the ``Weiss oscillations'' (0.1T $\leq B_0 \leq $ 1T), which at
higher fields (typically $B_0 \geq 0.6$T at $T=4.2$K) are superimposed by the
Shubnikov-de Haas oscillations. The resistivity component $\rho_{yy}$ does 
not exhibit such a positive magnetoresistance, but shows Weiss oscillations 
of opposite phase. No significant effect of the modulation on the Hall
resistance has been observed. Experiments on samples with a stronger electric
modulation \cite{Winkler89:1177,Beton90:9229},
and also the recent experiments on magnetically modulated systems
\cite{Carmona95:3009,Ye95:3013}, led to similar results, although
they usually did not provide the full information about the conductivity
tensor.  
The Weiss oscillations were first understood within a quantum mechanical
picture
\cite{Zhang90:12850,Peeters93:1466}:
The modulation lifts the degeneracy of the Landau levels and 
leads to dispersive Landau bands of oscillatory widths. The group velocity of
these bands leads to a ``band conductivity'', which explains the Weiss
oscillations of $\rho_{xx} $. The oscillatory density of states affects
 the scattering rate and gives rise to the Weiss oscillations of $\rho_{yy}$
\cite{Zhang90:12850,Pfannkuche92:12606}. 
Beenakker \cite{Beenakker89:2020} pointed out that the Weiss oscillations of
$\rho_{xx}$ can also be understood classically in 
terms of a guiding center drift of cyclotron orbits in the 
electric modulation field. He proved from Boltzmann's equation that, under
the assumption of a constant relaxation time, $\rho_{xx} $ is the only
component of the resistivity tensor that is affected by the modulation. The
low-$B_0$ positive magnetoresistance   has also been explained
classically \cite{Beton90:9229}, as being due to channeled trajectories which,
in $x$-direction are localized within a single period of the
modulation. Beenakker neglected these channeled orbits in his explicit 
calculation and  could therefore not explain the positive magnetoresistance,
whereas Beton {\em et al.}  \cite{Beton90:9229} could not extend their
calculation to the regime of Weiss oscillations. 
So far only  the semi-classical approach by
St\v{r}eda and coworkers \cite{Streda90:11892,Streda91:162,Kucera97:14439}
has been applied 
to both regimes.  However, this approach relies on quantum concepts such as
energy band structure and  magnetic breakdown.  A unified calculation, which
explains both the low-$B_0$ magnetoresistance and the Weiss oscillations of
$\rho_{xx}$ totally within classical dynamics, is still missing.   
The purpose of this paper is to present such a classical calculation.

We will extend Beenakker's work \cite{Beenakker89:2020} by including 
other modulation sources in
addition to a periodic electric field.  We will consider a magnetic modulation
and also a ``mobility modulation'', i.e. a  periodic position dependence of the
collision operator $C$ of Boltzmann's equation,  describing a 
spatial modulation of the density of scatterers. We will also allow for
anisotropic scattering of the electrons by, e.g., impurity potentials of finite
range, and thus go beyond the relaxation time approximation of Boltzmann's
equation. Within our 
classical approach we will prove that, in the absence of a 
mobility modulation, a one-dimensional superlattice defined by $x$-dependent
electric and magnetic fields affects only $\rho_{xx}$, even if anisotropic
scattering is included.  Allowing for a periodic mobility modulation  in
$x$-direction, we can obtain 
 oscillations of $\rho_{yy}$, however with the same 
phase as those of $\rho_{xx}$ in the case of a pure electric modulation.

After giving some details about our model in Sect.~II, we present
the formal calculations revealing the structure of the magnetoresistance
tensor in Sect. III and in Appendix \ref{appstructure}. In Sect.~IV we present
analytical results: For the regime of Weiss oscillations, we generalize
 Beenakker's calculation to include magnetic and mobility modulations,
and compare with simplified calculations based on the evaluation of the drift
velocities of cyclotron orbits
\cite{Beenakker89:2020,Gerhardts92:3449,Gerhardts96:11064}. 
For the regime of channeled trajectories we concentrate on the collisionless
limit, where we obtain  an algebraic dependence of the conductance
on the modulation strength. Therefore, in contrast to the regime of Weiss
oscillations, a simple power expansion of the transport coefficients with
respect to the modulation strength is not possible in this regime. 

Numerical results based on
a Fourier expansion of Boltzmann's equation are presented in Sect.~V. For
sufficiently weak modulation strengths, these results cover the whole range of
low-field magnetoresistance and Weiss oscillations, and they include the
effect of anisotropic scattering as well as of electric, magnetic and mobility
modulations. In a final Sect. \ref{summary} we summarize the most important
results. 

\section{The model}
It is known from quantum mechanical calculations
\cite{Zhang90:12850,Peeters93:1466} that the Weiss oscillations
on typical high-mobility samples are observed in a peculiar temperature range.
The temperature must be high enough, so that in the regime of Weiss
oscillations the Shubnikov-de Hass oscillations are not resolved. On the other
hand, it must be low enough, so that the transport coefficients are determined
only by the electronic states near the Fermi energy
\cite{Pfannkuche92:12606,Gerhardts92:3449,Gerhardts96:11064}. To describe the
relevant physics in this temperature window, 
we consider the 2DES in the $x$-$y$ plane as a degenerate Fermi gas, with Fermi
energy $E_F=(m/2)v_F^2$, of (non-interacting) particles with effective mass $m$
and charge $-e$ obeying classical dynamics. 
The velocity ${\bf v} =\dot{\bf r} = (\dot{x},\dot{y},0)$ 
of an electron obeys Newton's equation
\begin{equation}
m \dot{\bf v} = -e \left[ {\bf F} + ({\bf v} \times {\bf B})/c \right] \, .
\label{newton}
\end{equation}
To describe a modulated system with period $a$ in equilibrium, we write the
  electric field as ${\bf  F}(x)={\bf \nabla} V(x)/e$  and the magnetic field
  as   ${\bf B}(x) = (0,0, B_0 +B_m (x))$, and assume that the periodic
functions $V(x)=V(x+a)$ and $B_m(x)=B_m(x+a)$ have zero average values.
The thermal equilibrium at constant temperature $T$ and electrochemical
  potential $\mu^* $ is described by the distribution function
\begin{equation}
f_{eq}({\bf r},{\bf v})= f_0 (E({\bf r},{\bf v});T,\mu^*) \, ,
\label{fequilibrium}
\end{equation}
where the Fermi function $f_0(E;T,\mu^*)=1/[1+\exp [(E-\mu^*)/k_B T]] $ depends
on the dynamical variables only through the conserved energy 
\begin{equation}
E({\bf r},{\bf v}) =(m/2) {\bf v}^2 + V(x) \, .
\label{energy}
\end{equation}

In more general situations, the distribution function $f({\bf r},{\bf
  v},t) $, which yields  electron density $n_{el}$ and current density
  ${\bf j}$ according to 
\begin{equation}
n_{el}({\bf r},t)=\langle f \rangle _v \, , ~~~~ {\bf j}({\bf r},t)=\langle
-e{\bf v}f \rangle _v  \, ,
\label {dichten}
\end{equation}
where $\langle f \rangle _v \equiv [2m^2 /(2\pi \hbar)^2] \int d^2 v f$,
 is determined by Boltzmann's equation 
\begin{equation}
\left( \frac{\partial}{\partial t} + \dot{\bf r} \cdot
  \frac{\partial}{\partial  {\bf r} }+\dot{\bf v} \cdot
  \frac{\partial}{\partial  {\bf v} } \right) f = C[f; {\bf r},{\bf v},t]
\label{boltzmann}
\end{equation}
and suitable boundary conditions. For any reasonable microscopic model of
elastic and inelastic scattering \cite{Ziman:1960}, the collision operator
$C$ vanishes after averaging over the velocity,
$\langle C \rangle _v =0$. As a consequence, the $v$-integral over
Eq.~(\ref{boltzmann}) yields the continuity equation, $-e \partial n_{el} /
\partial t  +  {{\bf \nabla}\cdot{\bf j}} =0$. If we want to choose a
phenomenological model for $C$, we should be careful not to destroy this
property. 

In the following we consider the stationary, linear response of the modulated
system to an external homogeneous electric field ${\bf E}^0$, which
describes the average effect of a  voltage applied across the
sample. The result will be a non-zero current density ${\bf j}(x) $ and a
correction $\delta n_{el}(x)$ to the equilibrium electron density. At
low temperatures [$k_BT \ll \mu^* -V(x)$], the equilibrium density follows from
Eqs.~(\ref{fequilibrium}) and (\ref{dichten}) as $n_{el}(x)=D_0 [E_F -V(x)]$,
where $D_0=m/(\pi \hbar^2)$ is the density of states and $E_F
=\mu^*(T \rightarrow 0)$ the Fermi energy of the homogeneous 2DES.

With the ansatz 
\begin{equation}
f_{stat}({\bf r}, {\bf v}) = f_0 (E({\bf r},{\bf v});T,\mu^*)
-e \frac{\partial f_0}{\partial \mu ^*} \phi({\bf r},{\bf v}) 
\label{ansatz}
\end{equation}
for the stationary solution, and the assumption that
  the correction $\phi({\bf r},{\bf v})$ to the equilibrium  distribution is 
linear in ${\bf E}^0$, we obtain the linearized Boltzmann equation
\begin{equation}
{\cal D}\phi({\bf r},{\bf v}) - C[\phi;{\bf r},{\bf v}]={\bf v} \cdot {\bf
  E}^0 \, ,
\label{linearized}
\end{equation}
where ${\cal D} = {\bf v} \cdot
  {\partial}/{\partial  {\bf r} }+\dot{\bf v} \cdot
  {\partial}/{\partial  {\bf v} }$, and $\dot{\bf v}$ is determined by
  Eq.~(\ref{newton}) with the force in the equilibrium state. 
For $T  \rightarrow 0$, $\partial f_0 / \partial \mu^* \rightarrow \delta
  (E({\bf r},{\bf v})-E_F)$ and with Eq.~(\ref{energy}), which implies for the
  magnitude of the velocity $v(x)=v_F[1 -V(x)/E_F]^{1/2}$, the 2D $v$-integral
  in Eq.~(\ref{dichten}) reduces to an integral over the polar angle in the
  velocity space, e.g.
\begin{equation}
{\bf j} (x)=e^2 D_0 \int^{\pi}_{-\pi} \frac{d\varphi}{2\pi} v(x) {\bf
  u}(\varphi) \phi (x,\varphi) \, ,
\label{strom}
\end{equation}
where ${\bf u}(\varphi) =(\cos \varphi, \sin \varphi)$, and the dependence of
$\phi$ on $E_F$ is not explicitly indicated. The drift term of
Eq.~(\ref{linearized}) reads in these variables
\begin{equation}
{\cal D} = v(x) \cos \varphi \frac{\partial}{\partial x} +[\omega_c +
\omega_m (x)+  \omega_e(x,\varphi)] \frac{\partial}{\partial \varphi} \,
, \label{drift}
\end{equation}
with $\omega_c =e B_0 /mc$ the cyclotron frequency due to the average magnetic
field, and $\omega_m (x)= e B_m (x)/mc$ and $\omega_e(x,\varphi) = - \sin
\varphi \, dv/dx $ are due to the magnetic and the electric modulation,
respectively. 
The  collision operator describes the change of the distribution function due
to scattering events with a spatial extent which is short on the scale of the
drift motion \cite{Ziman:1960}. If there are different 
scattering mechanisms, we may write  $C[\phi;x,\varphi]= \sum_j
C_j[\phi;x,\varphi]$, with
\begin{equation}
C_j [\phi;x,\varphi]=  \frac{1}{\tau_j (x) }  \int
\frac{d\varphi'}{2\pi} P_j (\varphi' - \varphi)   [\phi(x,\varphi') -
\phi(x,\varphi)]  \, ,
\label{collision}
\end{equation}
where we have expressed the renormalized differential cross sections 
in terms of relaxation times $\tau_j (x)$ and dimensionless kernels
$P_j(\varphi )$.
Here, the last term describes the decay of the distribution function due to
scattering processes of the $j$-th type which start in phase space at
$(x,\varphi)$, whereas $\phi (x',\varphi ')$ describes the back-scattering into
this state. Assuming that the distribution function changes little on the
range of the scattering processes, we consider only $x'=x$ in the
back-scattering term. For scattering by impurities, the scattering
rate is proportional to the impurity density, which may be a function of the
position coordinate $x$ in periodically modulated samples. In order to
describe a periodic modulation of the electron mobility, we write
 $1/\tau _j (x) = 1/ \bar{\tau}_j + r_j (x)$ and
 assume that the oscillating parts $ r_j (x) =r_j(x+a)$
of the scattering rates have zero average values. Further we assume that the
kernels $P_j (\varphi)$, 
with the normalization $ \int^{\pi}_{-\pi} d\varphi P_j(\varphi) =2\pi$, have
the symmetry $P_j(-\varphi)=P_j(\varphi)$. This follows from the microscopic
symmetry if the effect of the magnetic field during an individual scattering
process is neglected.

It is interesting to note that the trigonometric functions are
eigenfunctions of the collision operator: for $\psi_m(\varphi)=\exp
(im\varphi)$ one has
\begin{equation}
C[\psi_m;x,\varphi]= -  [1 /\tau  ^{(m)}_{tr}  (x)] \,
\psi_m(\varphi) \, , \label{ceigen}
\end{equation}
with the eigenvalues
\begin{equation}
1/ \tau ^{(m)}_{tr} (x) =  \sum_j\, (1- \gamma_m^{(j)})/ {\tau_j (x)} \, ,
\label{tautr_m}
\end{equation}
where $\gamma_m^{(j)} = (2\pi )^{-1} \int^{\pi}_{-\pi}d\varphi P_j(\varphi)
\cos (m \varphi)$. The same holds for the real part and the imaginary part of
$\psi_m$ separately. We will demonstrate in the following
that several transport properties  do not depend on
all the details of the collision  operator defined by Eq.~(\ref{collision}),
but only on the transport time $\tau _{tr} (x) \equiv \tau ^{(1)}_{tr} (x)$,
which we write in the form $1/ \tau _{tr} (x) = 1/ \bar{\tau}_{tr} + r_{tr}(x)
$. For instance,  the Drude conductivity tensor of the homogeneous 2DES, with
$ r_{tr}(x) \equiv 0$, depends only on $\bar{\tau}_{tr}$. Thus, the transport
scattering rates due to the different scattering mechanisms simply add.

The special case 
$P_j(\varphi)\equiv 1$ describes {\em isotropic} scattering, and is identical
 with  the relaxation time approximation, $C_j[\phi ] =
-(\phi - \phi_{loc})/\tau_j$, where the relaxation is towards the
local equilibrium distribution $\phi_{loc} (x)= \int^{\pi}_{-\pi} d\varphi \,
\phi(x,\varphi)/2\pi$. This is the correct form of the relaxation time
approximation for inhomogeneous systems \cite{Mermin70:2362,Hu89:8468}.
 Neglecting $\phi_{loc}$ would describe relaxation towards the
total equilibrium distribution [cf. Eq.~(\ref{ansatz})], and would violate
explicitly the equation of 
continuity. As we will discuss below, this can lead to qualitatively wrong
results for the tranport coefficients.

\section{Structure of the resistivity tensor}
In this section we prove from Boltzmann's equation that a combined
electric and magnetic modulation in $x$-direction affects only the component
$\rho _{xx}$ of the effective resistivity tensor, if there is no
position dependence of the scattering rates, $r_j (x) \equiv 0$. 
All other components then
remain those of the unmodulated system. If, on the other hand, there is only a
modulation of the scattering rates while $V(x) \equiv 0$ and $B_m(x) \equiv
0$, only $\rho _{yy}$ will be affected. To obtain (within the classical
Boltzmann equation approach) a modulation effect on more than one component of
the resistivity trensor, one needs a mobility modulation in addition to an
electric and/or a magnetic modulation. We will make explicit use of the
equation of continuity to derive these results, which generalize an early
observation by Beenakker \cite{Beenakker89:2020} to more
general types of modulations and scattering mechanisms. Beenakker solved the
problem for a 
purely electric modulation in the relaxation time approximation. To appreciate
the important role of the continuity equation in this context, it is
instructive to consider first the simple local limit.

We want to emphasize that we assume translational
invariance in the $y$-direction throughout this work,
 so that, together with the modulations,
current densities and electric field components may depend only on $x$. 
Boundary effects are neglected. The experimental situation of a Hall bar with
 current flow in $x$-direction will be simulated by assuming
that the {\em average} current density in $y$-direction vanishes. The
experimental boundary conditions of exactly vanishing $j_y(x)$ on the sample
boundaries in $y$-direction would destroy the translational invariance in
$y$-direction and, thus, the simplification resulting from the merely
unidirectional modulation. They would require a
fully numerical treatment already in the local limit \cite{Jou96:203}, which
may become necessary under certain circumstances, but will not be discussed in
the present work.
\subsection{Local limit}  \label{locallimit}
In the local limit we describe the linear relation between current density
${\bf j}(x)$ and driving electric field ${\bf F}(x)$ by 
\begin{equation}
{\bf \hat{\rho}} \, ^{loc} (x) \, {\bf j}(x) =  {\bf F}(x) \, , \label{loclim}
\end{equation}
with a local resistivity tensor ${\bf \hat{\rho}} \, ^{loc} (x)$. It is
obtained from  the Drude resistivity tensor ${\bf \hat{\rho}}^D$ of the
homogeneous system, with components $\rho^D _{xx}=
\rho^D _{yy} = \rho_0 $ and $\rho^D_{xy}= - \rho^D_{yx}
= \omega_c \bar{\tau}_{tr}  \rho_0 $, where  $\rho_0  = m/( e^2
\bar{\tau}_{tr} \bar{n}_{el})$, by replacing the spatially constant electron
density $\bar{n}_{el}$, cyclotron frequency $\omega_c $, and transport time
$\bar{\tau}_{tr}$ by their spatially modulated counterparts 
$n_{el}(x)= \bar{n}_{el} [1-V(x)/E_F]$,  $\omega (x)=\omega_c +\omega_m (x)$,
 and $\tau_{tr} (x) = \bar{\tau}_{tr} /
[1+\bar{\tau}_{tr} r_{tr}(x) ]$, respectively. 
In thermal equilibrium, Eq.~(\ref{loclim}) applies with ${\bf j}(x) = 0$
and the vanishing
electrochemical field ${\bf F}(x)$  given by the  sum of the
electrostatic field ${\bf   \nabla} V(x) /e$ 
and the ``chemical'' contribution ${\bf \nabla} n_{el}(x) /(e D_0 )$.
The stationary response to a homogeneous external electric field
${\bf E}^0$ is accompanied by a change $\delta n_{el}(x)$ of the electron 
density, so that ${\bf F}(x) = {\bf E}^0 + {\bf \nabla} \delta n_{el}(x)
/(e D_0 )$ has to be inserted into  Eq.~(\ref{loclim}). 
Since $\delta n_{el}(x)$ must have the periodicity of the electric modulation,
we find $\langle {\bf F}(x) \rangle = {\bf E}^0 $,
where $\langle ...\rangle $ denotes the average over a period of the
modulation.
Since current density and field depend only on $x$, in the stationary state
the continuity equation $ {\bf \nabla} \cdot {\bf j} =0$ and Maxwell's
equation ${\bf \nabla} \times {\bf F} =0$ imply that $j_x$ and $F_y$ must be
independent of $x$. With this result, we can define an effective
resistivity tensor ${\bf \hat{\bar{\rho}}}$ by 
\begin{equation}
{\bf  \hat{\bar{\rho}}} \, \langle {\bf j}(x) \rangle  = {\bf E}^0 \, ,
\label{rho} 
\end{equation}
and calculate it by considering two experimentally relevant situations.
First, we assume that the average current density in $y$-direction is
zero. Then, solving the $y$ component of Eq.~(\ref{loclim}) for $j_y(x)$, we
obtain
\begin{equation}
 {\bar{\rho}}_{yx} = \frac{F_y}{  j_x} = \frac{ \langle
   {\rho}^{loc}_{yx} / {\rho}^{loc}_{yy} \rangle }{\langle
1/  {\rho}^{loc}_{yy} \rangle } \, . \label{locyx}
\end{equation}
 Inserting $ j_y(x)$ into the $x$-component of Eq.~(\ref{loclim}),
one obtains
\begin{equation}
 {\bar{\rho}}_{xx} = \frac{\langle F_x\rangle }{  j_x} = \left\langle
 {\rho}^{loc}_{xx} + [{\bar{\rho}}_{yx} -{\rho}^{loc}_{yx}] \,  \frac{
 {\rho}^{loc}_{xy} }{ {\rho}^{loc}_{yy}}  \right\rangle \, . \label{locxx}
\end{equation}
In the second, even simpler calculation, we set $j_x =0$ and obtain
\begin{equation}
 {\bar{\rho}}_{yy} = \frac{F_y}{ \langle j_y \rangle } = \frac{ 1}{ \langle 1/
 {\rho}^{loc}_{yy} \rangle } \,  \label{locyy} 
\end{equation}
and
\begin{equation}
 {\bar{\rho}}_{xy} = \frac{\langle F_x\rangle }{ \langle j_y \rangle }
=\frac{ \langle
   {\rho}^{loc}_{xy} / {\rho}^{loc}_{yy} \rangle }{\langle
1/  {\rho}^{loc}_{yy} \rangle }  \label{locxy} \, .
\end{equation}
Since ${\rho}^{loc}_{yx} =- {\rho}^{loc}_{xy}$, one finds from
Eqs.~(\ref{locyx}) 
and (\ref{locxy}) $ {\bar{\rho}}_{yx} =-  {\bar{\rho}}_{xy} $. If there is
no mobility modulation, $r(x) \equiv 0$, one has $\langle
1/  {\rho}^{loc}_{yy} \rangle = {\sigma}_0$ with ${\sigma}_0
 = 1/ \rho _0$ the Drude conductivity of the
 homogeneous  2DES.
Since $\langle \omega (x)\rangle  \bar{\tau}_{tr} = \omega _c \bar{\tau}_{tr}$,
Eqs.~(\ref{locyx}), (\ref{locxy}), and (\ref{locyy}) reduce to the
corresponding components of the Drude resistivity for the homogneous system.
Only $ {\bar{\rho}}_{xx} = \langle \rho ^{loc}_{xx} (x) \rangle +
\bar{\tau}_{tr}^2 [ 
\langle \omega (x) ^2 \rho_{xx}^{loc} (x) \rangle - \omega _c ^2 / 
{\sigma}_0] $ is different from the Drude value $1/{\sigma}_0 $, and shows a
positive magnetoresistance ($\propto \omega _c ^2$). Note that this result
differs from the simple average $\langle 
\rho^{loc}_{xx} \rangle$, which does not yield any magnetoresistance.
If only the mobility is modulated, one has $\langle
1/  {\rho}^{loc}_{yy} \rangle = {\sigma}_0 \, \langle \tau_{tr} (x) \rangle
/\bar{\tau}_{tr}$ and only $ {\bar{\rho}}_{yy}$ deviates from the
corresponding Drude value for the unmodulated system. If all types of
modulation are present, all components of the effective resistivity tensor
deviate from the Drude values. For weak modulations, one obtains up to second
order in the modulation amplitudes
\begin{eqnarray}
\frac{\bar{\rho}_{xx} - \rho ^D_{xx}}{\rho_0} &=& \left\langle
\left(\frac{V}{E_F} \right)^2 + \left( \omega_c \bar{\tau}_{tr} \frac{V}{E_F}
  +\bar{\tau}_{tr} \omega_m \right)^2 \right\rangle \nonumber \\
 &~&   \hspace*{1.6cm}     + \left\langle \bar{\tau}_{tr} r_{tr} \frac{V}{E_F}
\right\rangle \, , \label{loc2xx} \\
\frac{\bar{\rho}_{xy} - \rho ^D_{xy}}{\rho_0} &=&   -\, 
\left\langle \bar{\tau}_{tr} r_{tr} \left( \omega_c \bar{\tau}_{tr} \,
  \frac{V}{E_F}   
+\bar{\tau}_{tr} \omega_m \right) \right\rangle \, , \label{loc2xy} \\
\frac{\bar{\rho}_{yy} - \rho ^D_{yy}}{\rho_0} &=& -  \left\langle (
\bar{\tau}_{tr} r_{tr} )^2 \right\rangle - \left\langle  \bar{\tau}_{tr} r_{tr}
\frac{V}{E_F}  \right\rangle
\, . \label{loc2yy}
\end{eqnarray}

We want to emphasize that, in order to obtain these results, it was important
to satisfy the equation of continuity. If instead we would have taken naively
the spatial average of the local conductivity tensor, the resulting effective
resistivity tensor would not have the correct structure. A pure electric
modulation would have no effect at all. A pure magnetic modulation would
affect $\bar{\rho}_{xx}$ and  $\bar{\rho}_{yy}$ in the same manner. Although
the leading order correction to the $xx$-component would agree with
Eq.~(\ref{loc2xx}), the corresponding correction to the $yy$-component would
disagree with the correct result (\ref{loc2yy}).

\subsection{Nonlocal calculation}
The local limit applies to systems in which the electronic mean free path is
much shorter than the modulation period. In order to exhibit the Weiss
oscillations, the system must be in the opposite limit, with a mean free path
considerably larger than the modulation period.  
For such systems, we  consider  the current density (\ref{strom})
resulting from the solution of Boltzmann's equation (\ref{linearized}) and
define the effective resistivity tensor by Eq.~(\ref{rho}).
We want to emphasize that the spatial average affects only $j_y (x)$, whereas
$j_x$ must be independent of $x$ by virtue of the continuity equation.

By formal, but exact manipulations of the Boltzmann equation we show in
Appendix \ref{appstructure} that the tensor inverse of  $ {\bf
  \hat{\bar{\rho}}} $ can be written as
\begin{equation}
{\bf \hat{\bar{\sigma}}} = {\bf \hat{\sigma}}^D  + \rho_0 \, {\bf \hat{\sigma}}^D 
{\bf \hat{\Gamma}} \, {\bf \hat{\sigma}}^D \, \label{sigmahat}
\end{equation}
where ${\bf \hat{\sigma}}^D$ is the Drude conductivity tensor of the
corresponding homogeneous system, and  ${\bf \hat{\Gamma}}$ has the components
\begin{eqnarray}
\Gamma_{xx} &=& - \langle \bar{\tau}_{tr} r_{tr} V/E_F \rangle - G_{xx} \, , ~~
\Gamma_{xy} = - G_{xy} \, \nonumber \\
 \Gamma_{yy} &=& + \langle \bar{\tau}_{tr} r_{tr} V/E_F \, \rangle + G_{yy} ,
 ~~ \Gamma_{yx} = + G_{yx} \, , \label{Gamma}
\end{eqnarray}
with
\begin{equation}
G_{\mu \nu} = \frac{2 \bar{\tau}_{tr}}{ v_F^2} \left\langle  \int_{-\pi}^{\pi}
\frac{d \varphi}{2 \pi} \, g_{\mu} \chi_{\nu} \right\rangle \, . \label{G_munu}
\end{equation}
Here 
\begin{equation}
g_x = \frac{v_F^2}{2}  \frac{dV/dx}{E_F} +  \omega_m v_y \, , ~~
g_y = - r_{tr} v_y \, , \label{g_xy}
\end{equation}
and $ \chi_{\nu}$ is the solution of the modified Boltzmann equation
\begin{equation}
[{\cal D} -C ] \, \chi _{\nu} = g_{\nu} \, . \label{chi_nu}
\end{equation}
Then, by tensor inversion,  $ {\bf \hat{\bar{\sigma}}}^{-1} = {\bf
  \hat{\bar{\rho}}} = {\bf \hat{\rho}}^D + \Delta {\bf \hat{\rho}}$, we obtain from
  Eq.~(\ref{sigmahat})
\begin{equation}
\Delta {\bf \hat{\rho}} / \rho _0 = - {\bf \hat{\Gamma}} \left[ {\bf \hat{1}}
+ \rho _0 \, {\bf \hat{\sigma}}^D {\bf \hat{\Gamma}} \right] ^{-1} \,
. \label{delrho} 
\end{equation}
Explicit matrix inversion and multiplication finally leads to the relatively
simple result
\begin{equation}
\frac{\Delta {\bf \hat{\rho}}}{ \rho _0 } = -\left[\,  {\bf \hat{\Gamma}}
 +\frac{{\rm det}
 {\bf \hat{\Gamma}} }{1+ \omega_c ^2 \, \bar{\tau}_{tr}^2 } \,  \frac{ {\bf
 \hat{\rho}}^D }{ \rho _0 } \, \right] \, \frac{1}{D} \, , \label{drhofin}
\end{equation}
with ${\rm det} {\bf \hat{\Gamma}}  =
\Gamma_{xx}\Gamma_{yy}-\Gamma_{xy}\Gamma_{yx}$ and
\begin{equation}
D=1+\frac{\Gamma_{xx}+\Gamma_{yy}+\omega_c  \bar{\tau}_{tr} \left(
\Gamma_{xy}-\Gamma_{yx} \right) + {\rm det} {\bf \hat{\Gamma}}  }{ 1+
 \omega_c ^2 \, \bar{\tau}_{tr}^2 } \, . \label{determinante}
\end{equation}
If $\Gamma_{xx}$ or $\Gamma_{yy}$ is the only nonzero matrix element of ${\bf
  \hat{\Gamma}}$, Eq.~(\ref{drhofin}) reduces to (\ref{delrhoxx}) or
  (\ref{delrhoyy}), respectively, i.e., either only $\Delta \rho _{xx}$ or
  only $\Delta \rho _{yy}$ is nonzero, and we obtain the same tensor structure
  as in the local limit. As shown in Appendix
  \ref{appstructure}, these results can be derived without the explicit formal
  result (\ref{drhofin}).

We want to emphasize that only the global transport time $\bar{\tau}_{tr}$,
defined below Eq.~(\ref{tautr_m}), enters these
considerations about the structure of the effective resistivity tensor, not
any further details of the scattering mechanisms. The latter enter, of
course, the solutions of Eq.~(\ref{chi_nu}) and determine, according to
(\ref{G_munu}), the numerical values of the tensor components of ${\bf
  \hat{\Gamma}}$.

\section{Analytic results}
\subsection{Weiss oscillations}  \label{Weiss_osc}
If the modulation is weak, in the sense that the forces due to the electric and
the magnetic modulations are weak as compared to the Lorentz force due to the
average magnetic field $B_0$, a perturbation expansion with respect to the
amplitudes of the modulation is possible. To obtain the conductivity correct
to second order, we need, according to Eq.~(\ref{G_munu}), $\chi_{\nu}$ to
first order, and we can replace the square bracket in Eq.~(\ref{delrho}) by
unity. Since the right hand side of Eq.~(\ref{chi_nu}) is already of first
order, we should neglect modulation effects on $C$ and on ${\cal D}$,
i.e. replace in Eq.~(\ref{drift}) $v(x)$ with $v_F$ and omit $\omega_m(x) $ and
$\omega_e (x,\varphi)$. 
In this approximation the different Fourier coefficients of the expansions
$V(x)=\sum_{q\neq 0}V_q \exp (iqx)$, and similar for $ \omega_m(x)$ and $
r_{tr}(x)$ with coefficients $\omega_q$ and $r_q$, respectively,
do not mix in Eq.~(\ref{chi_nu}) and lead to uncoupled integro-differential
equations of the variable $\varphi$.  We now follow Beenakker
\cite{Beenakker89:2020} and consider only isotropic scattering, i.e. we take
$\gamma_0^{(j)} =1$ and $\gamma_m^{(j)} =0$ for $m\neq 0$ in
Eqs.~(\ref{ceigen}), and define the relaxation time $\tau \equiv \bar{\tau }$
by the spatial average $1/ \tau = \langle 1/ \tau (x) \rangle$. Then we can
solve these equations analytically and obtain
\begin{eqnarray}
G_{xx} &=& \frac{1}{2} \sum_{q\neq 0}\,  
 \left\{ | {\lambda qV_q}/{E_F} | ^2 Q_q^{ee}    \right. \label{G_xx}  \\
 &~& ~ -   \left.  [4\lambda q \tau \omega_q V_{-q}/{ E_F}] \,Q_q^{em}   
 +   |2 \tau \omega_q|^2 Q_q^{mm} \right\} \, , \nonumber \\
  G_{xy}  &=&  \sum _{q \neq 0} \tau r_{-q}  [(\lambda q V_q /E_F)
Q_q^{em} -2\tau \omega_q Q_q^{mm}] \, , \label{deltaxy} \\
G_{yx} &=& -\, G_{xy}  \, , ~~~  ~    G_{yy} =  2 \sum_{q \neq 0}   |\tau r_q
|^2 Q_q^{mm} \, , \label{G_yy} 
\end{eqnarray}
with $Q_q^{ee}= S_q^{(0,0)}/N_q$, $Q_q^{em}= S_q^{(0,1)}/N_q$, and $Q_q^{mm}=
S_q^{(1,1)} + (S^{(0,1)}_q) ^2/N_q$, where $N_q=1-S_q^{(0,0)}$ and
\begin{equation}
S^{(\mu,\nu)}_q = \sum_{m=-\infty}^{\infty}
\frac{J_m^{(\mu)}(Rq) \, J_m^{(\nu)}(Rq)}{1+(m\omega_c\tau)^2} \, ,
\label{Smunu}
\end{equation}
where $\lambda = v_F \tau$ is the mean free path,
$R=v_F /\omega _c$ is the cyclotron radius, and $J_m^{(0)}(z)=J_m (z)$
and $J_m ^{(1)}(z) =J'_m (z)$ are the Bessel functions and their derivatives.
For $\omega_q \equiv 0$, $r_q\equiv 0$,  and $V_q =\delta
_{|q|,2\pi/a}V_{rms}/\sqrt{2}$ this 
reduces to Beenakker's result \cite{Beenakker89:2020}. The term $S_q^{(0,0)}$
in the denominator $N_q$ arises from the back-scattering term in the collision
operator, which describes relaxation towards the local equilibrium. It is thus
a consequence of the equation of continuity.

\subsubsection{Clean limit, $\omega_c \tau \gg 1$} \label{cleanlimit}
In the limit $\omega_c
\tau \gg 1$, only the term with $m=0$ survives in the sum of Eq.~(\ref{Smunu}),
and in Eq.~(\ref{delrhoxx}) the term $\tilde{\Gamma}_{xx}$ may 
be neglected, so that $\Delta \rho_{xx} / \rho_0 \approx G_{xx}$. Then,
rearranging terms with $q$ and $-q$, the curly bracket in 
Eq.~(\ref{G_xx})  can be replaced by $|\lambda q {\cal S}_q /E_F |^2$ with
${\cal S}_q = V_q J_0(Rq)+(k_F/q)\hbar \omega_q J_1(Rq)$,
\begin{equation}
\Delta \rho_{xx} / \rho_0 \approx \frac{1}{2}  \sum_{q\neq 0}\,  \frac{ |\lambda q {\cal S}_q /E_F |^2 }{ 1- J_0^2(Rq) } \, . \label{classic}
\end{equation}
 This result is, apart from the
denominator $1-J_0^2(Rq)$,  in agreement with a recent simplified calculation
\cite{Gerhardts96:11064} based on an evaluation of Chambers' formula for the
conductivity in terms of the drift velocity of cyclotron orbits.  That
calculation 
corresponds to a naive relaxation time approximation in Boltzmann's equation
with relaxation towards the total instead of the
local equilibrium distribution function, and thus violates the equation of
continuity for the modulated system. Indeed, the nontrivial denominator in the
present calculation results from the back-scattering term in
Eq.~(\ref{collision}). 

In Fig.~\ref{beenakker} we demonstrate the range of validity of several
approximations to the results (\ref{G_xx}) - (\ref{Smunu})
for the special case of a simply harmonic electrical modulation $V(x)=V_0 \cos
qx$. Figure~\ref{beenakker}(a) shows, for several values of the
mean free path $\lambda$, the quantity $S_q^{(0,0)}$ of Eq.~(\ref{Smunu}),
which determines the 
denominator of the magnetoresistance corrections. Figures~\ref{beenakker}(b)
and (c) show results for the magnetoresistance, calculated from
Eq.~(\ref{G_xx}),  with different normalizations 
which are suitable to discuss the local limit, (b), and the limit of infinite
 $\lambda$, (c), respectively. The curves are plotted versus the inverse
 cyclotron radius, which is proportional to the average magnetic field $B_0$.
In the limit $\lambda \rightarrow \infty $,
 $S_q^{(0,0)}=J_0^2(Rq) \approx (2/\pi Rq) \cos ^2 (Rq- \pi /4)$ 
decreases
with decreasing $B_0$, with an infinite number of zeroes and with values at the
maxima, which approach zero linearly with $B_0$. The same then holds for the
magnetoresistivity curve obtained from Eq.~(\ref{classic}) and for the thin
line in Fig.~\ref{beenakker}~(c), which corresponds to the simplified result of
Refs. \cite{Gerhardts92:3449,Gerhardts96:11064} and is 
calculated from Eq.~(\ref{classic}) by replacing the denominator by 1. 
For $1/Rq <0.1$ the denominator of Eq.~(\ref{classic})
is close to unity ($S_q^{(0,0)}< 0.1$), and the thin line approximates the
result of Eq.~(\ref{classic}) quite well. At
higher magnetic fields, $1/Rq > 0.2$, the result of Eq.~(\ref{classic})
agrees well with the magnetoresistivity curve obtained for $\lambda q= 30$ in
Fig.~\ref{beenakker}~(c). For a sample with  a given value of the mean free
path, Eq.~(\ref{classic}) does not hold uniformly for all values of the average
magnetic field. With decreasing $B_0$ values, $\lambda q/ Rq = \omega_c
\tau $ becomes smaller and more and more terms contribute to the sum in
Eq.~(\ref{Smunu}). As a consequence, the minima of the Weiss oscillations of 
 $S_q^{(0,0)}$ and $\Delta \rho _{xx}$ are risen to positive values, the
 amplitudes 
 of the oscillations are increasingly suppressed with decreasing $B_0$ values,
 and, in the limit $B_0 \rightarrow 0$,  a finite positive value is approached
 with zero slope. 

Whereas $S_q^{(0,0)}$ in the limit
of large $B_0$ ($Rq \rightarrow 0$)
approaches 1  for all values of $\lambda$ [since $J_0(0)=1$ and $J_m(0)=0$ for
$m\neq 0$], it shows an overall increase with
decreasing $\lambda$ for all values of $Rq$, and approaches 1 uniformly in
$B_0$ for $\lambda q \rightarrow 0$ [since 
$\sum_{-\infty}^{\infty} J_m^2(z) \equiv 1$]. Thus, the
denominator $N_q$ in Eqs.~(\ref{G_xx})-(\ref{G_yy}) becomes most important in
the local limit $\lambda q \rightarrow 0$ and in the ``quasi-local'' limit
$Rq<1$, where the cyclotron radius $R$ becomes smaller than the period $a=2\pi
/q$ of the modulation. 
Figure \ref{beenakker}~(b) shows the magnetoresistivity data for $\lambda q=9$
and $\lambda q=4$ in a representation which is more convenient for the
discussion of the local limit $\lambda q \rightarrow 0$. For $\lambda q=1$,
the corresponding result approximates already closely the local result
of Eq.~(\ref{locxx}),  $\Delta
\rho _{xx} /\rho _0 =(1/2)(V_0/E_F)^2 \, [1+ (\lambda / R)^2]$, which is
indicated by the dash-dotted line.

In the ``quasi-local'' high magnetic field limit, $Rq \ll 1$, the denominator
becomes   $1-J_0^2(Rq) \approx (Rq)^2/2 \ll 1$, so that, for  $V(x)=V_0 \cos
qx$, we obtain a quadratic magnetoresistance, 
\begin{equation}
\Delta \rho _{xx} / \rho_0 \approx (\omega_c \tau)^2
(V_0/E_F)^2/2 \, , ~~~ (Rq \ll 1) \, , \label{largeBel}
\end{equation}
 whereas the  calculation of
Ref.\cite{Gerhardts96:11064} yields a saturation for large $\omega_c$, $\Delta
\rho _{xx}^{[15]} 
/ \rho_0 \approx (\lambda
q)^2(V_0/E_F)^2/4$. Experiments seem 
to show a quadratic magnetoresistance at high magnetic fields, although masked
by the onset of Shubnikov-de Haas oscillations. For a sinusoidal magnetic
modulation, $\omega_m (x) = \omega_m^0 \cos qx$, we obtain from
Eq.~(\ref{classic}) at large $B_0$ values a saturation,
\begin{equation}
\Delta \rho _{xx} / \rho_0 \approx (\omega_m^0 \tau )^2 /2 \, , ~~~
(Rq \ll 1) \, , \label{largeBma}
\end{equation}
whereas Ref.\cite{Gerhardts96:11064} yields $\Delta
\rho _{xx}^{[15]}
 / \rho_0 \approx (\lambda
q)^2 (\omega_m^0 /\omega_c )^2 /4$, which becomes small $\propto B_0^{-2}$. We
want to emphasize that, in this ``quasi-local'' limit, the $B_0$ dependence
of the magnetoresistance obtained from Eq.~(\ref{classic})
agrees with that of the local approximation (\ref{loc2xx}).

Figure \ref{beenakker2} shows the prediction of Eqs.~(\ref{G_xx}) -
(\ref{Smunu}) for the case when two types of modulations are simultaneously
present. In Fig.~\ref{beenakker2}~(a) an electric and a magnetic cosine
modulation of comparable effective strengths are considered. The dash-dotted
line refers to a pure electric  and the solid line to a pure magnetic
modulation, whereas the dashed lines refer to a combination of both. For fixed
amplitudes of the electric and the magnetic cosine modulations, the resulting
magnetoresistivity curve depends strongly on the phase difference between
both. The long-dashed line is Fig.~\ref{beenakker2}~(a) is obtained for
in-phase modulations, and the short-dashed line for a phase shift of a half
period. A systematic investigation of the interference effects occurring due
to the superpositon of an electric and a magnetic modulation has been given 
within the simplified treatment of  Ref.\cite{Gerhardts96:11064}. Since for 
$\lambda q > 20$  this treatment is qualitatively correct in the regime of
Weiss oscillations, we will not repeat this discussion here.

Figure \ref{beenakker2}~(b) shows the components of the magnetoresistivity
tensor for a combination of electric and mobility modulations.  We assume here
in-phase cosine
modulations for the electrostatic potential energy and the scattering rate of
the electrons, and comment on opposite phases below. According to
Eqs.~(\ref{G_xx}) -  (\ref{Smunu}), $\Delta 
\rho _{xx}$ is determined by the quantity $Q_q^{ee}$, $\Delta\rho _{yy}$ by 
$Q_q^{mm}$, and $\Delta\rho _{xy} = - \Delta\rho _{yx}$ by $Q_q^{em}$, 
all of which contain different combinations of Bessel functions and their
derivatives. For $\lambda q > 20$, in the regime of Weiss oscillations
$Q_q^{ee}$ is proportional to the square of $J_0 (Rq)$ and $Q_q^{mm}$, which
also determines the results of a pure magnetic modulation, is proportional to
the 
square of the derivative $J'_0 (Rq)$. This leads to  antiphase oscillations
of $Q_q^{ee}$ and   $Q_q^{mm}$ as  depicted in Fig.~\ref{beenakker2}~(a) by
the dash-dotted and the solid lines, respectively. However, according to
Eqs.~(\ref{Gamma}), (\ref{delrho}), and (\ref{G_xx}) - (\ref{G_yy}),
 $\Delta \rho _{xx}$ and $\Delta\rho
_{yy}$ exhibit {\em in-phase oscillations} since,
apart from $B_0$-independent offsets $+ \tau r_m^0 (V_0/E_F)/2$ and 
 $- \tau r_m^0 (V_0/E_F)/2$, respectively,
$\Delta \rho _{xx}$ is proportional to  $Q_q^{ee}$, whereas $\Delta\rho _{yy}$
is proportional to  $ - Q_q^{mm}$. Thus, a mobility modulation can not explain
the {\em anti-phase oscillations} of $\Delta\rho _{yy}$ observed in the early
experiments \cite{Weiss89:179}, as mentioned in the Introduction. In the
situation considered in Fig.~\ref{beenakker2}, $Q_q^{em}$ is approximately
given by the derivative of $Q_q^{ee}$ with respect to $Rq$. Thus in 
 Fig.~\ref{beenakker2}~(b), where the resistivity components are plotted versus
$1/Rq$, $\Delta\rho _{yx}$ appears as the derivative of $\Delta\rho _{xx}$. If
we change the relative sign of electric and mobility modulation,  the sign of
$\Delta\rho _{yx}$ will also change, whereas the phase relation between
$\Delta \rho _{xx}$ and $\Delta\rho _{yy}$ remains unchanged.

We do not show a figure analog to  Fig.~\ref{beenakker2}~(b) for a
superposition of a magnetic and a mobility modulation, since in this case all
 components of the magnetoresistivity tensor are determined by   $Q_q^{mm}$. As
 a consequence, $\Delta\rho _{yx}$ would show the same appearance as
 $\Delta\rho _{xx}$, i.e. as the solid line in Fig.~\ref{beenakker2}~(a), and 
$\Delta\rho _{yy}$ would appear as $-\Delta\rho _{xx}$, i.e. with the opposite
phase. In contrast to the case of electric and mobility modulation, in the
case of magnetic and mobility modulation there is no constant offset in the
diagonal components of the resistivity tensor, and  $\Delta\rho _{yx}$ does not
assume  negative values.

\subsubsection{Damping of Weiss oscillations}
The value of $S_q^{(0,0)}$ for weak magnetic fields, $Rq \geq 2$, depends
strongly on $\lambda q$, and Weiss oscillations are washed out for $1/Rq <
0.5/ \lambda q$, i.e., for $\omega_c \tau < 0.5$. If the mean free path is too
small, $\lambda q \leq 2$ (here we have in
mind that the period of the modulations, and thus $q$ is fixed), no
oscillations survive, since too many Bessel functions contribute to the sum in 
Eq.~(\ref{Smunu}) ($\omega_c \tau < 1$ in the $1/Rq$ regime where oscillations
can be expected).  In this regime the calculation of
Ref.\cite{Gerhardts96:11064} is not applicable, since it assumes $\lambda \gg
R$.  

We can also re-obtain the local limit $\lambda q \rightarrow 0 $
from Eqs.~(\ref{G_xx}) - (\ref{G_yy}), if we expand the Bessel functions in
Eq.~(\ref{Smunu}) for small arguments, $Rq \rightarrow 0 $, and keep only the
leading orders. This means that we consider the limit of large modulation
period, but make no assumption about the magnitude of $\omega_c \tau$.
 In this limit, only the terms with $m=0$ and $m=\pm 1$
contribute to Eq.~(\ref{Smunu}), and we exactly recover Eqs.~(\ref{loc2xx}) -
(\ref{loc2yy}), with $\tau$ instead of $\bar{\tau}_{tr}$. To obtain this
result correctly, it is essential to calculate the denominator  (which
vanishes in this limit) consistently up to the order $(Rq)^2$, which includes
contributions from the Bessel functions $J_0$ and $J_1$. This demonstrates
nicely that the denominator in   Eqs.~(\ref{G_xx}) - (\ref{G_yy}) indeed
is important in order to obtain the correct local-limit results,
Eqs.~(\ref{loc2xx}) - (\ref{loc2yy}), which were derived by exploiting
explicitly  the equation of continuity.

\subsection{Low-field magnetoresistance }
If the average magnetic field $B_0$ becomes too small, the analytic solution of
Eq.~(\ref{chi_nu}) for small modulation amplitudes is no longer applicable. It
was obtained under the assumption, that for the calculation of $\chi_{\nu}$ 
 the modulation contributions $\omega_m(x)$ and
$\omega_e(x,\varphi)$ to the drift operator (\ref{drift}) could be neglected as
compared to the average cyclotron frequency $\omega_c$. Physically this means
that the modulations modify the cyclotron motion in the average magnetic field
only weakly and lead essentially to a drift of the cyclotron orbits. If
however, for a given modulation, $B_0$ becomes smaller than a critical
$B_{crit}$ (or $R > R_{crit}$), this is no 
longer true and other types of orbits (``channeled orbits'') occur which are
confined in the $x$-direction to a single period of the modulation
superlattice. The effect of these orbits, which lead to a positive
magnetoresistance at small $B_0$ \cite{Beton90:9229}, is not included in
Eqs.~(\ref{G_xx}) - (\ref{G_yy}). 

\subsubsection{Classification of orbits}
To characterize the channeled orbits and to elucidate their importance for the
magnetoresistance,  we proceed as follows. We note that, for a
one-dimensional modulation in $x$-direction, there exists   a second conserved
quantity in addition to the energy, Eq.~(\ref{energy}).  Integrating the
$y$-component of Newton's equation (\ref{newton}), we can write
\begin{equation}
v_y -\frac{e}{mc} \left[ xB_0 + \int_0^x d x' B_m(x') \right] = \frac{p_y}{m}
\, , \label{p_y}
\end{equation}
where $p_y$ is a constant of motion, which has the meaning of a canonical
momentum.  
Solving for $v_y$ and inserting this into Eq.~(\ref{energy}), we obtain for
the projection of the orbit on the $x$-direction 
\begin{equation}
(m/2) v_x^2 + \tilde{V} (x) = E_F \, , \label{eff1d}
\end{equation}
with the effective one-dimensional potential
\begin{equation}
\tilde{V}(x)=V(x) + \frac{m}{2} \omega _c ^2 \left[ x - x_0 + \int_0^x d x'
\frac{B_m(x')}{B_0} \right]^2 \, . \label{V_eff}
\end{equation}
Here we assume $B_0 \neq 0$, and the constant of motion is written as
 $x_0 =-c p_y /e B_0$, which in the absence of  modulations is the $x$
 coordinate of the center of the cyclotron orbit. For each value of $x_0$ one
 obtains a different location of the effective potential
 $\tilde{V}(x)$  on the $x$-axis, and thus a different orbit located between 
turning points, at which $v_x =0$. If $x$ is a turning point, $\tilde{V}(x) =
 E_F$, Eq.~(\ref{V_eff}) yields two possible values for the constant $x_0$:
\begin{equation}
x_0^{\pm}(x) = x +\int_0^x d x' \frac{B_m(x')} {B_0 } \, \pm \, R \,
\sqrt{1-\frac{V(x)}{E_F}} \, .
\label{x0pm}
\end{equation}
Since $\tilde{V}(x) \leq E_F $ holds for any orbit passing through a given
position $x$, only orbits with  
 $x_0$ values in the interval $x^-_0(x) \leq x_0 \leq x^+ _0(x)$ can pass
 through $x$. (We specify orbits only up to an arbitrary shift in $y$
 direction.) On the other hand, a given value of $x_0$ determines an orbit
 between turning points $x_t$ satisfying $x_0^- (x_t) = x_0$ or $x_0^+ (x_t) =
 x_0$. Thus the curves $ x_0 = x_0^{\pm}(x)$ yield the constant of motion of
 the orbits which have $x$ as a turning point. We use this property to
 distinguish two types of orbits. 

Orbits of the first type we call ``drifting
 orbits''. These have their left turning point $x_l$ satifying $x_0^+ (x_l) =
 x_0$  and their right turning point $x_r$ satifying $x_0^- (x_r) = x_0$. For
 sufficiently smooth modulation (cf. Appendix B) and large $B_0$ this is the
 only type of orbits, as we demonstrate in the left panels of Fig.~\ref{orbits}
 for an electric cosine modulation. In the limit of vanishing modulation these
 ``drifting orbits'' reduce to cyclotron orbits, i.e., to circles, and $x_r -
 x_l \rightarrow 2 R$. The effect of the modulation in $x$-direction is to
 deform the circular cyclotron orbits and to superimpose on the cyclotron
 motion a drift velocity in $y$-direction. The drift vecolity depends on the
 position of the orbit in the modulation field and may vanish for
 highly symmetric positions. In the left panel of Fig.~\ref{orbits} one orbit
 is nearly symmetric with respect to the potential maximum at $x=0$, another
 one nearly symmetric about the minimum at $x=1.5 a$. The small distance
 between  neighboring loops of the orbits (three loops are shown) 
 indicates a low drift velocity. 

Orbits of the second type, which we call ``channeled orbits'', occur if the
curves $x_0 ^{\pm} (x)$ have local extrema, i.e., for sufficiently strong
modulation or weak magnetic field $B_0$. For ``channeled orbits'' we require
that their turning points both satisfy either $x_0^+ (x_l) =x_0^+ (x_r) = x_0$
or $x_0^- (x_l) =x_0^- (x_r) = x_0$, i.e. in plots like those  in the upper
panels  of Fig.~\ref{orbits} both turning points are located either on $x_0^+
(x)$ or on  $x_0^- (x)$. These channeled orbits (see middle and right panels of
Fig.~\ref{orbits}) are confined to a single
period $a$ of the modulation and have no counterparts in the unmodulated
system. For an electric modulation they occur near  potential minima. For a
magnetic modulation they occur where the total magnetic field $B_0
+ B_m (x)$ changes sign \cite{Mueller92:385}.

Each value $x_0$ of the constant of motion determines uniquely a single
drifting orbit (of course, modulo a shift in $y$-direction). Channeled orbits
may occur with the same $x_0$ value, however only in intervals of the $x$ axis
which are not entered by that drifting orbit. At a given position $x$, each
allowed $x_0$ value characterizes uniquely a single orbit through $x$.
 In the limit $B_0 = 0$, drifting orbits degenerate into wavy trajectories
 with periodic velocity ${\bf v}(x+a) = {\bf v}(x)$ and nonzero mean velocity
$\langle v_x \rangle $ in $x$-direction. The mean value in $y$-direction is
$\langle v_y \rangle = p_y /m$ (note that $x_0 /R = - p_y /m v_F$, and 
for pure electric modulation $ v_y  = p_y /m$ is constant).
 In this limit two trajectories with 
$\langle v_x \rangle > 0 $ and  $\langle v_x \rangle < 0 $ occur with the same
$ \langle v_y \rangle $, as indicated by the dashed lines in the lower right
panel of Fig.~\ref{orbits}.

If $x_0$ coincides with the value of a local maximum of $x_0^- (x)$ or a local
minimum of $x_0^+ (x)$, the orbits near the corresponding turning point $x_t$
become critical in the sense that they do not reach this turning point, but
approach in the $x-y$ plane the straight line $x=x_t$ asymptotically.

\subsubsection{Effect of channeled orbits}
As has been emphasized by Beton et {\em al.} \cite{Beton90:9229}, the
channeled orbits are responsible for the strong positive magnetoresistance of
modulated systems in weak perpendicular magnetic fields.
In Appendix \ref{appb_0} we calculate
the fraction $f_{co}(B_0,V_0)$ of the phase space covered by channeled
orbits, as a function of the (average) magnetic field $B_0$ and  the modulation
strengths $V_0$,  for two electrostatic modulation models. For the smooth
cosine modulation with a fixed $V_0$, channeled orbits exist only at
sufficiently low $B_0 < B_{crit}(V_0)$, whereas for a 
discontinuous step modulation skipping orbits survive at arbitrarily high
magnetic fields.  For both models the fraction of channeled orbits at zero
magnetic field, $f_{co}(0,V_0)$, increases at low modulation strengths
proportional to  $|V_0 /E_F |^{1/2}$.   We also sketch, 
for  $B_0=0$ and in the nonlocal limit of large mean free path, a calculation
in the spirit of Ref.~\cite{Gerhardts96:11064} which shows
 that the resistivity correction  increases as $|V_0 /E_F |^{3/2}$ with the
 modulation amplitude. This indicates that, for $B_0=0$ and weak modulation, no
 expansion of the transport coefficients in a power series of $V_0$ is
 possible. The same is true at finite magnetic field if channeled orbits
 exist. Indeed, for the step model which allows for channeled orbits at arbitrary
 strong $B_0$, the sum over the Fourier coefficients in the second-order formula
 (\ref{G_xx}) for the magnetoconductivity diverges for all values of $B_0$,
 indicating that such an expansion does not exist \cite{Gerhardts92:3449}.

To include the effect of channeled orbits on the magnetoresistance properly, we
have solved Boltzmann's equation numerically. The results  show in the limit
of long mean free path and zero magnetic 
field the same $|V_0/E_F |^{3/2}$ behavior as obtained analytically in the
simplified approach, and will be presented in the next section.

\section{Numerical results}
\subsection{Some technical details}
In order to cover the whole magnetic field range, we have solved
Eq.~(\ref{chi_nu}) numerically, using  Fourier expansions for the dependences
on both $x$ and $\varphi$.   According to Eq.~(\ref{ceigen}), 
 the collision operator, Eq.~(\ref{collision}), is diagonal in the Fourier
 representation. The drift term, Eq.~(\ref{drift}), becomes simple for strictly
 harmonic modulations, provided that the  electric modulation is weak enough
 and $v(x)$ can be approximated by $v_F$. Then the drift term couples only 
  neighboring Fourier coefficients, and Eq.~(\ref{chi_nu}) leads to an infinite
  set of
  linear equations for the Fourier coefficients with a sparsely occupied
  matrix. Truncating the infinite set by considering only the lowest $N_x$ and
  $N_{\varphi}$   Fourier coefficients for the $x$- and the
  $\varphi$-dependence, respectively, we could exploit the sparse nature of
  the system matrix and handle approximations of the infinite set with large
  dimensions $ N_x \cdot  N_{\varphi}$. The number of Fourier coefficients
  needed 
  to obtain converged results increases with increasing modulation strenghts
  and with increasing mean free path. The convergence in the regime of Weiss
  oscillations is much better than in the low-$B_0$ regime, where
  the channeled orbits are important. For the pure electric cosine modulation
  with $V_0/E_F =0.02$ we needed $N_x=13$ and $N_{\varphi}=550$ to achieve
  convergence in the low-magnetic-field regime $5 \cdot 10^{-5} < 1/qR < 0.02$
  , whereas   $N_x=4$ and $N_{\varphi}=450$ was sufficient for $1/qR >
  0.5$. The reason for the poor convergence  at low $B_0$ is that the
  distribution function   calculated from  Eq.~(\ref{chi_nu}) developes  very
  sharp and rapidly changing structures near the angles $\varphi = \pi /2$ and
  $3 \pi /2$, originating from the turning points of the channeled
  orbits. Owing   to these numerical problems, our Fourier expansion approach
  is reliable only   for weak modulations ($V_0/E_F \leq 0.2$).
\subsection{Relaxation time approximation}
  A typical low-magnetic-field result for the pure electric modulation
  $V(x)= V_0 \cos qx$ with $V_0 /E_F =0.02$ is shown  in
Fig.~\ref{fig5_11}.  The solid line is the result of our numerical
calculation. The dotted line shows the result of the analytical approximation
(\ref{G_xx}). The dash-dotted line indicates the contribution of the channeled
orbits and is obtained as follows \cite{Beton90:9229}. For weak modulation and
$\omega _c \tau \gg 1$, the correction to the Drude resistivity tensor can be
written as $\Delta \rho _{xx} /\rho _0 \approx  \omega _c^2 \tau ^2 \Delta
\sigma _{yy} / \sigma _0$. The contribution of channeled orbits may be
  estimated by their mean square drift velocity 
$\Delta \sigma _{yy} / \sigma _0 = \langle v_d^2 \rangle /(v_F^2/2)$,
cf. Eq.~(\ref{v_v_corr}). Approximating $v_d^2 \approx v_F^2$, this becomes $
  \approx 2 f_{co}^{\rm cos} (B_0,V_0)$, i.e.,
 twice the fraction of channeled orbits at the Fermi energy (see appendix
 \ref{appb_0}). The  dash-dotted line in Fig.~\ref{fig5_11} represents 
$\Delta \rho _{xx} /\rho _0 \approx 2 \omega _c^2 \tau ^2 f_{co}^{\rm cos}
(B_0,V_0)$ with an additional vertical shift to match our numerical result at
$B_0=0$. The vertical
  straight line indicates the critical magnetic field 
  $B_{crit}^{Beton} =cq V_0 /e v_F$ at which the channeled orbits die out
  \cite{Beton90:9229}.  Our numerical   result   exhibits a maximum 
  at the magnetic field value $B_p  \approx 0.8 
  B_{crit}^{Beton}$. This is in very good agreement with recent experiments
  \cite{Soibel97:4482}, although the numerical factor 0.8 should depend on the
  specific form of the potential modulation. 
The corresponding   results for a pure magnetic modulation look qualitatively
  very similar and   are not shown. 

The inset of Fig.~\ref{fig5_11} shows our numerical results
  for strong purely magnetic modulations of the form $B_m (x)= B_m^0 \cos qx$.
  The modulation strength is given in units of $1/qR_m =aeB_m^0 /(2\pi m c
  v_F)$, $\Delta \rho _{xx}/ \rho _0$ in units of $\omega _m^2 \tau ^2 =
  \lambda ^2 / R_m ^2$. 
With increasing modulation strength (different curves in the inset of
  Fig.~\ref{fig5_11}), the extent of the positive
  magnetoresistance regime at low $B_0$ values increases and $\Delta \rho
  _{xx}$ reaches a maximum
  at $B_p \approx 0.7 B_m^0 $ (note that $B_{crit}=B_m^0$, see appendix
  \ref{appb_0}). The magnetoresistivity at $B_p$ increases with increasing
  modulation strength, and the low-field Weiss oscillations are successively
  suppressed 
  (the fluctuations in the curve for $1/qR_m =0.34$ are due to numerical
  errors, which become larger with increasing modulation strength). Again the
  dependence of the magnetoresistance on the modulation strength is very
  similar for the electric cosine modulation and will not be shown
  explicitly. For large mean free path  and $V_0/E_F   > 0.2$, fluctuations of
  the calculated curves indicate convergence 
  problems. We found, however, that the relation $B_p   \approx 0.8
  B_{crit}^{Beton}$, i.e. a linear increase of $B_p $ with the modulation
  amplitude $V_0$, is reasonably well satisfied up to $V_0/E_F \approx
  0.5$, in agreement with Ref.~ \cite{Beton90:9229} and with a recent
  modification of St\v{r}eda's semiclassical approach \cite{Kucera97:14439}.
The peak    values of the   magnetoresistivity  can be well 
  fitted by 
\begin{equation}
\Delta \rho_{xx}(B_p)/\rho_0 =7.01\, (V_0/E_F)^2+54.7 \, (V_0/E_F)^3 \,
. \label{Vhoch3}
\end{equation}
Whereas in the regime of Weiss oscillations ($B_0 > 3 B_{crit}$) $\Delta
\rho_{xx}$ is proportional to $(V_0/E_F)^2$ and the analytic approximation
(\ref{G_xx}) is valid, the peak value at $B_p$ is dominated by the third order
term in Eq.~(\ref{Vhoch3}) for $V_0/E_F >0.15$. 
For weak modulation ($V_0/E_F < 0.15$), $\lambda q \gg 1$ and $B_0=0$, our
numerical results are consistent with the analytically calculated algebraic
dependence $\Delta \rho_{xx}(0)/\rho_0 \propto (V_0/E_F)^{3/2}$ (see
Appendix \ref{appb_0}). This indicates that, at low magnetic fields where
channeled orbits 
exist, an expansion of the magnetoresistance in powers of the modulation
strength is not possible. To check this, we expanded the Boltzmann
equation (\ref{chi_nu}) in powers of $V_0/E_F$ and solved it up to the order $
 (V_0/E_F)^8$. In the regime around $B_p$ the higher order contributions lead
 to high-frequency oscillations, but we found  no indication of
 convergence or positive magnetoresistance.

\subsection{Anisotropic scattering} \label{anisotropic}
Experiments in the regime of Weiss oscillations give evidence that the
mobility of the 2DES is dominated by scattering of the electrons by remote
ionized impurities, which should lead to predominant forward scattering. To
investigate the effect of anisotropic scattering, we simulate the angular
dependence of the differential
scattering cross section in the collision operator, Eq.~(\ref{collision}),  by
the simple model
\begin{equation}
P_p(\varphi) = c + (1-c)\, \frac{(p!)^2}{(2p)!} \left(2 \cos \frac{\varphi}{2}
\right)^{2p} \, , \label{Pforward}
\end{equation}
where $c$ controls a constant background ($0 \leq c \leq 1$), and $p$ the
 sharpness of the forward scattering peak. This model is convenient, since it
 has simple Fourier coefficients, $\gamma _0 =1$, $\gamma _m =(1-c) (p!)^2
 /[(p+m)! (p-m)!]$ for $|m| \leq p$, and $\gamma _m =0$ for $|m| >p$
 [cf. Eq.~(\ref{ceigen})]. According to Eq.~(\ref{tautr_m}), the ratio of
 transport time and relaxation time $\tau$ increases  then  with increasing
 $p$ as $\tau _{tr} / \tau = (p+1)/(cp+1)$. 

Keeping $\tau$ fixed, we have calculated magnetoresistivity curves for several
values 
of the parameters $c$ and $p$.  We found as a general trend that, with
inreasing $\tau _{tr} / \tau $, for both electric and magnetic modulations the
values of $\Delta \rho _{xx} / \rho_0$ [with $ \rho_0 =m/(e^2 \bar{n}_{el}
\tau _{tr})$] increase slightly faster than proportional to  $\tau _{tr} /
\tau $. On an absolute scale this means that the values of $\Delta \rho _{xx}
/ \rho_0$ change little near well developed minima but increase stronly at the
peaks, resulting in a strong increase of the amplitudes of the Weiss
oscillations. In Fig.~\ref{fig5_35} we have, for increasing $\tau _{tr}$,
reduced the relaxation time $\tau$ so that the maximum value of $\Delta \rho
_{xx} / \rho_0$ near $1/qR =0.26$ remains unchanged. In this plot higher-index
Weiss oscillations are increasingly damped with increasing importance of
small-angle scattering. Comparison with typical experiments shows that 
small-angle scattering is indeed important. Apart from the quantum mechanical
Shubnikov-de Haas oscillations at large $B_0$, we could nicely fit
experimental curves in the whole region $1/qR \leq 0.8$ with 
typical  values of $\tau _{tr} / \tau \geq 2$.
\subsection{Mobility modulation}
Finally, we have considered, within the relaxation time approximation, a mixed
electric modulation $V(x)=V_0 \cos qx$ and mobility modulation $1/ \tau (x) =
\bar{\tau}^{-1} + r_m \cos qx$. We choose an in-phase modulations, $V_0 \cdot
r_m >0$. We found that, for $V_0/E_F \ll 1$ and $\bar{\tau} v_F q \gg 1$, the
effect of the mobility modulation on $\rho _{xx}$ is rather small.
This is understandable in the regime of Weiss oscillations, since,
according to Fig.~\ref{beenakker}, $\Delta \rho _{xx} / \rho _0$ becomes
independent of $\lambda q$  for $\lambda q \geq 25 $. 
 It is also expected for $B_0=0$
in the local limit, where $\Delta \rho _{xx}/ \rho _0$ becomes proportional
to $(V_0/E_F)\cdot \bar{\tau} r_m$ (For the corresponding step modulations one
has $\bar{ \rho} _{xx}/ \rho _0 = [1+ (V_0/E_F) \bar{\tau} r_m ]/[1 -
(V_0/E_F)^2]$.) 

In Fig.~\ref{fig5_45} we show the results for pure electric and pure mobility
modulations as solid lines. Whereas the electric modulation enhances $\rho
_{xx}$, the mobility modulation reduces $\rho _{yy}$. The Weiss oscillations
of both resistivity components are in phase, as we have seen already from the
analytical solution Eq.~(\ref{G_xx}) - (\ref{G_yy}). For the pure mobility
modulation, there is no positive magnetoresistance at small $B_0$ values,
since there are no channeled orbits in this case.

Results for the case of
mixed modulations are shown as broken lines. For the diagonal resistivities,
the changes compared to the pure cases are essentially a positive offset of
$\rho _{xx}$ and a negative offset of $\rho _{yy}$, as expected from
Eqs.~(\ref{Gamma}) and (\ref{drhofin}). The enhancement of $\rho_{xx}$ at small
$B_0$ values ($R > R_{crit}$) is plausible, since now in the regime of
channeled trajectories the mobility is increased. Therefore, the importance of
these trajectories and thus the value of the magnetoresistivity is enhanced.

For the mixed case, we obtain also a correction to the Hall resistivity. To
accommodate this correction in the same figure, we have subtracted the Drude
value, which increases linearly with $B_0$. In the regime of Weiss
oscillations, $\Delta \rho _{xy}$ is roughly proportional to the slope of
$\rho_{xx} /\rho_0$ as function of $B_0 \propto 1/qR$.

\section{Summary}  \label{summary}
Using the linearized stationary Boltzmann equation and including anisotropic
scattering in the collision operator, we have investigated the effect of
one-dimensional periodic modulations on the resistivity tensor of a
two-dimensional electron system in a perpendicular magnetic field $B_0$. Even
in the relaxation time approximation (for isotropic scattering) 
we have carefully considered the back-scattering term of the collision
operator, which ensures relaxation of the distribution function towards its
local equilibrium value in the stationary state, and not to the total
equilibrium distribution. This is of principle importance for the modulated
systems, since it ensures that the approximation does not violate the equation
of continuity, and it has qualitative as well as quantitative consequences for
the resistivity tensor.

The qualitative consequences concern the form of the resistivity tensor. In the
absence of mobility modulations, electric and magnetic modulations in
$x$-direction affect only  $\rho _{xx}$, whereas the other components of the
resistivity tensor remain those of the unmodulated system. This result, which
is not obvious in the presence of a magnetic field, has first been obtained by
Beenakker \cite{Beenakker89:2020} for a pure electric modulation in the
relaxation time approximation. In Appendix A we show that this result also
holds for mixed electric and magnetic modulations and for anisotropic
scattering. Moreover, we show that, in the absence of electric and magnetic
modulations, a mobility modulation in $x$-direction affects only  $\rho _{yy}$.
In Sect.~\ref{locallimit} we have derived the same tensor structure in the
local limit. This more transparent derivation elucidates  the important role
played by the continuity equation.

Simpler calculations based on an evaluation of the familiar
velocity-velocity-correlation formula for the conductivity tensor are not in
agreement with these exact results on the structure of the resistivity tensor
\cite{Beenakker89:2020,Gerhardts92:3449,Gerhardts96:11064}, although they may
yield reasonable approximate results under certain conditions.

For weak modulations, isotropic scattering, and sufficiently high magnetic
field $B_0$, we followed 
an approximation scheme proposed by  Beenakker \cite{Beenakker89:2020} and
obtained analytic results for the magnetoresistivity tensor
(Sect. \ref{Weiss_osc}). In the clean limit (mean free path much larger than
modulation period) these results reduce to those of the simpler calculations,
apart from a characteristic denominator, which can be traced back to result
from the back-scattering term in the collision operator. The absence of this
denominator in the simplified treatment
\cite{Gerhardts92:3449,Gerhardts96:11064} indicates that this treatment does
not correctly describe the relaxation towards the local equilibrium, and thus
violates the equation of continuity. Quantitatively, the deviation of the
denominator from unity is not large if the cyclotron radius $R$ (of an
electron at the Fermi energy) is larger than the period $a$ of the
modulation. Therefore, in the regime of Weiss oscillations the simplified
treatment gives reasonable results. However, in the quasi-local regime of
high magnetic fields ($R<a$), the denominator becomes small and the simplified
treatment becomes wrong. In this regime the correct treatment of the equation
of continuity is quantitatively important, see Sect. \ref{cleanlimit}.

The analytic approximation, which implicitly assumes weakly perturbed
cyclotron motion, breaks down at low magnetic fields, where ``channeled
orbits'' become important \cite{Beton90:9229}.  
At $B_0=0$ and for $\lambda q \gg 1$, these should lead to an interesting
algebraic dependence of the resistivity on the modulation strength, as we
demonstrate in Appendix B.
To include the effect of the channeled orbits, we have solved the linearized
Boltzmann equation numerically. The 
numerical results in the relaxation time approximation (isotropic scattering)
give a qualitatively correct description of the experimental result for
$\rho_{xx}$ in the range of small magnetic fields, where a pronounced positive
magnetoresistance is obtained, and of intermediate magnetic fields, where the
Weiss oscillations are observed. At $B_0=0$, they are also consistent with
analytical results which are derived in Appendix B. A nearly quantitative
description of the experimental $\rho_{xx}$ \cite{Weiss89:179}
in the weak and intermediate magnetic field regime  is obtained
from a numerical calculation considering predominant forward scattering (Sect.
\ref{anisotropic}).

In an attempt to find a classical explanation of the {\em anti-phase} Weiss
oscillations observed in $\rho_{yy}$ \cite{Weiss89:179}, we have also
investigated a mobility modulation. We obtained, however, only weak {\em
  in-phase} 
oscillations, both analytically and numerically.
Thus, this attempt of a classical explanation of the $\rho_{yy}$  oscillations
fails, whereas the quantum treatment provides a
convincing explanation of these oscillations  as a
density-of-states effect  \cite{Zhang90:12850}.

\acknowledgments{ We gratefully acknowledge many fruitful and stimulating
  discussions with D. Pfannkuche and D. Weiss. We also thank U. Gossmann,
  G. Nachtwei, D. Pfannkuche, and D. Weiss for the careful reading of the
  manuscript and constructive critique.
This work was supported by the BMBF grant 01 BM 622. }

\appendix
\section{Moments of Boltzmann's equation} \label{appstructure}
To investigate the structure of the tensor ${\bf \hat{\rho}}$, we multiply the
linearized Eq.~(\ref{linearized}) with $e^2 D_0 v(x) 
{\bf u}(\varphi)$ and average with respect to both $ \varphi$ and $x$ over a
full period. 
Integrating by parts eliminates the derivatives of the distribution function
$\phi (x,\varphi )$,  see Eq.~(\ref{drift}), and the collision operator is
evaluated using the real and imaginary part of
\begin{equation}
\int^{\pi}_{-\pi} \frac{d \varphi}{2 \pi}\, e^{i\varphi} C[\phi; x, \varphi] 
= - \frac{1}{\tau_{tr} (x)} \int^{\pi}_{-\pi} \frac{d \varphi}{2 \pi}\,
e^{i\varphi} \phi (x,\varphi) \, , \label{evalcol} 
\end{equation}
where only the effective transport scattering rate
$1/\tau_{tr} (x)$ enters [cf. Eq.~(\ref{ceigen})]. Multiplying with
$\bar{\tau}_{tr}$ and rearranging terms we can write the result as
\begin{equation}
{\bf \hat{\rho}}^D \, \langle {\bf j}(x) \rangle  = 
{\bf E}^0 + {\bf \Delta} \, ,   \label{ohmeff}
\end{equation}
where ${\bf \hat{\rho}}^D$ is the Drude resistivity tensor of the homogeneous
system, and [with $V'(x)=dV/dx$]
\begin{eqnarray}
  \Delta_x &=& - \langle [V'(x)/E_F] \phi^{loc} (x) + \rho_0 
\bar{\tau}_{tr} \omega_m (x) j_y (x) \rangle \, , \label{delta_x}\\
 \Delta_y &=& -\rho_0 \bar{\tau}_{tr} \left\langle r_{tr} (x) 
 j_y (x) \right\rangle \, . \label{delta_y}
\end{eqnarray}
 Here we have explicitly used that,
according to the equation of continuity, $j_x (x)= \langle j_x (x) \rangle $
is spatially constant, and that the spatial average values of $ \omega_m (x)$,
$V(x)$, and $r_{tr}(x)$ vanish. 

For the unmodulated system ${\bf \Delta} = {\bf 0}$, and Eq.~(\ref{ohmeff})
reduces to the well known Drude form of Ohm's law with homogeneous current
density ${\bf j}(x) = \langle {\bf j}(x) \rangle ={\bf j}^0 $. For the
modulated system 
with the same ${\bf E}^0$, we may write $\langle {\bf j}(x) \rangle  = {\bf
  j}^0 + \langle \delta {\bf j}(x) \rangle $, so that
\begin{equation}
{\bf \hat{\rho}}^D \, \langle \delta {\bf j}(x) \rangle  =  {\bf \Delta} \, .
\label{rhodj} 
\end{equation}
 To calculate these corrections, it is convenient to write the solution of
 Eq.~(\ref{linearized}) in the form
\begin{equation}
\phi (x, \varphi ) =\phi ^0 (\varphi ) v(x)/v_F + \rho_0  \tau_{tr} \chi (x,
\varphi ) \, , 
\label{chiansatz}
\end{equation}
where $\phi ^0 (\varphi ) = \rho _0 \tau _{tr} v_F {\bf u} (\varphi ) \cdot
{\bf  j}^0$ is the corresponding solution for the unmodulated system. This
yields
\begin{equation}
 {\bf j}(x) = {\bf j}^0 v^2(x)/v_F ^2 + \delta {\bf j}(x) \, ,
 \label{j0deltaj}
\end{equation}
with
\begin{equation}
\delta {\bf j}(x) = \frac{2}{v_F^2} \int_{-\pi}^{\pi} \frac{d \varphi}{2\pi}
 \, v(x) {\bf u}(\varphi) \chi (x,\varphi ) \, , \label{deltaj}
\end{equation}
and $\chi
(x, \varphi )$ satisfies Eq.~(\ref{linearized}) with a modified inhomogeneity
which is linear in the modulation quantities  $ \omega_m (x)$,
$V(x)$, and $r_{tr}(x)$,
\begin{eqnarray}
[{\cal D} -C ] \, \chi = (&-& vdv/dx+\omega_m v_y - r_{tr} v_x ) j_x^0
\nonumber  \\  &-& (\omega_m v_x + r_{tr} v_y ) j_y^0  \, , \label{chi}
\end{eqnarray}
where $(v_x,v_y)=v(x)(\cos \varphi , \sin \varphi )$.
We may write $\chi = \chi_1 + \chi _2$, where $\chi _2$ solves 
\begin{equation}
[{\cal D} -C ] \, \chi _2 = 
- v(x) \cos \varphi \,[ r_{tr}  j_x^0+ \omega_m j_y^0 ] \, . \label{chi_2}
\end{equation}
The solution is independent of $\varphi$, and satisfies
\begin{equation}
d \chi _2 /dx = - [  r_{tr}  j_x^0+ \omega_m j_y^0 ] \, . \label{dchi_2}
\end{equation}
It  does not contribute to the current density,
Eq.~(\ref{deltaj}), but it contributes to the local equilibrium distribution,
and thus contributes to the inhomogeneity of Eq.~(\ref{delta_x}),
\begin{eqnarray}
\left\langle \frac{dV/dx}{E_F} \, \phi^{loc} \right\rangle &=& \rho_0
  \bar{\tau}_{tr} \, \left\langle \frac{dV/dx}{E_F} 
 \int_{-\pi}^{\pi} \frac{d \varphi}{2\pi} \,  \chi _1
  \right. \nonumber \\  &~& ~~~~~ \left.   
+ \frac{V}{E_F} \,
[  r_{tr}  j_x^0+ \omega_m j_y^0 ]\right\rangle \, . \label{evalchi_2}
\end{eqnarray}
From Eqs.~(\ref{j0deltaj}) to (\ref{evalchi_2}) we obtain
\begin{eqnarray}
\Delta _x =&-& \rho _0 \bar{\tau}_{tr} \left\langle r_{tr}  \frac{ V}{E_F}
 j_x^0  \right. \label{delta1_x} \\ 
     &~&    ~~  \left.
+ \frac{2}{v_F^2} \int_{-\pi}^{\pi} \frac{d \varphi}{2 \pi} \left(
\frac{v_F^2}{2}  \frac{dV/dx}{E_F} +  \omega_m v_y \right) \chi_1
 \right\rangle  \, , \nonumber \\
\Delta _y =&-& \rho _0 \bar{\tau}_{tr} \left\langle -  r_{tr}  \frac{ V}{E_F}
 j_y^0     + \frac{2}{v_F^2} \int_{-\pi}^{\pi} \frac{d \varphi}{2
 \pi} \, r_{tr}v_y   \chi_1 
 \right\rangle   . \label{delta1_y}
\end{eqnarray}
Finally, we write $\chi _1 = \chi _x j^0_x + \chi_y j^0_y$, where, according
to Eqs.~(\ref{chi}) and (\ref{chi_2}), $\chi_{\nu}$ (for ${\nu}=x,y$)
satisfies Eq.~(\ref{chi_nu}) with $ g_{\nu}$ given by Eq.~(\ref{g_xy}).
We obtain
\begin{equation}
{\bf \Delta} = \rho_0 {\bf \hat{\Gamma}} \, {\bf j}^0 \, ,\label{dgj0}
\end{equation}
where $ {\bf \hat{\Gamma}}$ is defined by Eq.~(\ref{Gamma}).

From these results we can show that the magnetoresistivity tensor reveals a
very simple structure in the two important special cases, where there is either
no modulation of the scattering rate, or where only the scattering rate is
modulated. Let us first assume that we have an electrical and a magnetic
modulation, but no mobility modulation, $r(x) \equiv 0$. 
Then $\Gamma_{xx}=- G_{xx}$ and all other tensor components of ${\bf
  \hat{\Gamma}}$ vanish, so that $\Delta_x =\rho_0
\Gamma_{xx} j^0_x$, while $\Delta_y =0 $. Thus, according to
Eq.~(\ref{rhodj}),  $\langle \delta j_y \rangle
= \omega_c \bar{\tau}_{tr}\langle \delta j_x \rangle $ and  
$\langle \delta j_x \rangle  = \tilde{\Gamma}_{xx} j^0_x$, with
$\tilde{\Gamma}_{xx} = 
\Gamma_{xx} /[1+(\omega_c \bar{\tau}_{tr})^2]$.
 If we now consider the effective conductivity
tensor ${\bf \hat{\sigma}} = {\bf \hat{\rho}}^{-1}$ of the modulated system,
${\bf j}^0 + \langle 
\delta {\bf j} \rangle ={\bf \hat{\sigma}} {\bf E}^0 $, we obtain for the
components of  ${\bf \hat{\sigma}}$ in terms of those of the Drude  ${\bf
  \hat{\sigma}}^D = [{\bf \hat{\rho}}^D]^{-1}$: $\sigma_{\mu \nu} =
\sigma_{\mu \nu}^D (1+ \tilde{\Gamma}_{xx})$ for $(\mu \nu)=(xx)$, $(xy)$,
and $(yx)$, whereas $\sigma_{yy}=\sigma_{yy}^D [1-(\omega_c
\bar{\tau}_{tr})^2\tilde{\Gamma}_{xx}] $. Tensor inversion shows that  ${\bf
  \hat{\rho}} $ differs from the Drude resistivity tensor
${\bf   \hat{\rho}}^D$ of the unmodulated system only in the $xx$ component,
\begin{equation}
{\Delta \rho _{xx}}/{ \rho_0} ={\rho _{xx}}/{ \rho_0} -1 = -
{\Gamma_{xx}}/[1+\tilde{\Gamma}_{xx}] \, , 
\label{delrhoxx}
\end{equation}
while all other components are not modified by the modulation. This result,
first pointed out by Beenakker \cite{Beenakker89:2020} for purely
electrostatic modulation within the relaxation time approach, is thus shown to
hold also in the presence of a 
magnetic modulation and  anisotropic scattering.

The other simple case is that without electric and magnetic modulation, $V(x)
\equiv 0$, $\omega_m(x) \equiv 0$, but with a finite mobility modulation
$r_{tr}(x)$. In this case $ \Delta_x =0$ and $ \Delta_y = \rho_0 \Gamma
_{yy} j^0_y$.  The same procedure yields now a resistivity tensor, which
differs from that of the unmodulated system only in the $yy$ component,
\begin{equation}
\Delta \rho _{yy} / \rho_0 = - \Gamma_{yy}/(1+ \Gamma_{yy}/[1 +(\omega_c
\bar{\tau}_{tr})^2])   \, . 
\label{delrhoyy}
\end{equation}

In the general case, with mobility modulation and electric or magnetic
modulation, we obtain from Eqs.~(\ref{rho}), (\ref{ohmeff}), and (\ref{dgj0})
the effective conductivity tensor given by Eq.~(\ref{sigmahat}).

\section{Channeled orbits} \label{appb_0}
\subsection{Critical field $B_{crit}$ }
For a smooth modulation, the
 condition for the existence of  channeled orbits is apparently, that the
 functions $x^-_0 (x)$ and  $x^+_0 (x)$ have local extrema, i.e., that the
 equation 
\begin{equation}
1+\frac{B_m(x)} {B_0 } \mp \frac{R}{2} \frac{V'(x)/E_F}{\sqrt{1-V(x)/E_F}} =0
\label{channeledcond}
\end{equation}
has a solution. For a pure magnetic modulation of the form $B_m(x)=B_m^0 \cos
qx$, we obviously find $B_{crit} = B_m^0$. For $B_0 <B_{crit}$ the total
magnetic field $B(x)=B_0 +B_m (x)$ changes sign, and channeled orbits exist
close to lines with $B(x)=0$ \cite{Mueller92:385}. 

For a pure electric modulation of
the form $V(x)=- V_0 \cos qx$, Eq.~(\ref{channeledcond}) is equivalent with a
quadratic equation for $ \cos qx $, which has a solution only if $1-(qR)^2 +
[(qR)^2 V_0 / 2 E_F]^2 >0 $. This yields for the critical cyclotron radius 
%
\begin{equation}
\frac{2}{qR_{crit}} = \frac{V_0}{E_F} \, \left[
\frac{2}{1+\sqrt{1-(V_0/E_F)^2}} \right] ^{ \frac{1}{2}} \, . \label{beton}
\end{equation}
For $V_0 \ll E_F$,  we recover the result $B_{crit} \approx
B_{crit}^{Beton}(V_0) =c qV_0 /e v_F$, which has previously been obtained by
Beton {\it et al.}  
\cite{Beton90:9229} from the balance of the average Lorentz force and the
maximum electrical force $qV_0$ due to the modulation.
For stronger modulation, the $B_{crit}$ obtained from Eq.~(\ref{beton})
increases faster than linearly with $V_0$ and reaches the value  $B_{crit} =
\sqrt{2} \, B_{crit}^{Beton}$ at $V_0 = E_F$. Then all trajectories are
confined to a single period of the potential. For $B_0 > B_{crit}$, i.e., for
$2R<a \sqrt{2}/ \pi $, all trajectories are ``drifting'' orbits, 
with left turning points on  $x^+_0 (x)$ and 
right turning points on $x^-_0 (x)$, whereas for $B_0 < B_{crit}$ there exist
also ``channeled'' trajectories having all their turning points either on
$x^+_0 (x)$ or on $x^-_0 (x)$.

For the step modulation $V(x)= - V_0 {\rm sign} (\cos qx )$, $x_0^+ (x)$ and
$x_0^- (x)$ are piecewise linear functions with discontinuities at the walls of
the potential wells, so that Eq.~(\ref{channeledcond}) can not be used to
determine $B_{crit}$. Two types of channeled orbits may exist
near a local minimum of $x_0^- (x)$ (maximum 
of $x_0^+ (x)$): traversing trajectories with turning points on
opposite walls of a potential well, and skipping orbits which are reflected
only by one of the walls. Traversing trajectories exist for sufficiently weak
magnetic fields, $B_0 < B_{crit}^{\rm step}(V_0) =mcv_F/eR _{crit}^{\rm step}$
with 
\begin{equation}
\frac{a}{2R _{crit}^{\rm step}} =
\sqrt{1+V_0/E_F} - \sqrt{1-V_0/E_F} \, . \label{critstep}
\end{equation}
Skipping orbits survive at arbitrarily high magnetic fields. Physically, this
is a consequence of the fact that the singular electric field at a potential
discontinuity never becomes negligible as compared to the Lorentz force due
to $B_0$. For small modulation amplitude we find $ B_{crit}^{\rm step}(V_0) =
(2/\pi )B^{Beton}(V_0)$. This indicates that the linear dependence of the
critical magnetic field on the (weak) modulation strength holds independently
of the detailed shape of the potential 
modulation, which may, however, affect the numerical prefactor.

A characteristic feature of the channeled orbits is that, even at $B_0 \neq 0$,
their velocity in $y$-direction, $v_y = \omega _c (x- x_0)$, does not change
sign. 
\subsection{Number of channeled orbits}
We now calculate the fraction of electrons occupying channeled orbits. The
electron density $n_{el}(x;E_F)= D_0 [E_F - V(x) ] \theta (E_F - V(x))$, i.e.,
the area density of occupied states at $x$ with energy below $E_F$, is
proportional to the area $\pi v^2(x)$ of the circle with radius $v(x)=v_F
[1-V(x)/E_F] $ in  ${\bf v}$-space. The areal density of states at $x$ with
energy at $E$ is 
given by the local density of states $D(x;E) = d n_{el}(x;E)/dE =D_0$. Both,
$n_{el}(x;E)$ and $D(x;E)$ are independent of the magnetic field $B_0$, and so
are  their spatial average values $\bar{n}_{el} (E)$ and $D(E)$. An electron
trajectory through $x$, at $B_0=0$, is a channeled orbit, if the energy
available for the 
motion in the $x$-direction is smaller than the height of the potential
barrier, $E_F - (m/2)v^2_y < V_{max}$, where $ V_{max}$ is the maximum value
of $V(x)$.  With $v_y =v(x) \sin \varphi$, this means $\sin ^2 \varphi > \sin
^2 \varphi_0 (x;E_F)$,  where
\begin{equation}
\varphi_0 (x;E_F) =\arcsin \sqrt{[E_F-V_{max}]/[E_F-V(x)]} \,
. \label{phi_xEF}  
\end{equation}
Thus, the contribution of channeled orbits to the local density of states is
$D_{co} (x;E) = D_0 [1- (2/\pi)  \varphi_0 (x;E)]$, and the density of
electrons in channeled orbits (with energy below $E_F$) is $n_{co}(x;E_F) =  
D_0 [E_F - V(x) ] \theta (E_F - V(x))$ if $E_F < V_{max}$, and 
\begin{eqnarray}
&~& n_{co}(x;E_F) =  \frac{2D_0}{\pi} \left[ \sqrt{E_F -V_{max}}\,
  \sqrt{V_{max}   -V(x)} \right. \nonumber \\ 
&~& ~~~~~~~~~~+\left. \left(E_F -V(x) \right) \arcsin \sqrt{\frac{V_{max}
  -V(x)}{E_F   -V(x) } }\, \right] \, , \label{nso_xEF}
\end{eqnarray}
if $E_F > V_{max}$.
For fixed average electron density $\bar{n}_{el}$, the Fermi energy is
independent of the modulation strength, if $V_0 < E_F = \bar{n}_{el} /D_0$.
For $V_0 > E_F$ and $B_0=0$, all trajectories become channeled orbits. For the
step potential $V(x)=-V_0 {\rm sign} (\cos qx)$, channeled orbits exist only in
the potential wells, and Eq.~(\ref {nso_xEF}) yields the area density of channeled
orbits in the well with $V_{max}=V_0$ and $V(x)=-V_0$. For $V_0 >E_F$ the
requirement of fixed average density $\bar{n}_{el}$ yields a constant value
of $E_F + V_0 = 2\bar{n}_{el} /D_0$ in the wells, so that for $V_0 > E_F$ the
situation is identical to that at $V_0 = E_F$, with vanishing electron density
in the barrier regions. Since the local density of states is $D_0$, we
normalize, for arbitrary $V_0$, the fraction of channeled orbits for the step
modulation on the total 
density of states in the wells, $ f^{\rm step}_{co}(B_0=0,V_0) = D^{\rm
  step}_{co} (x;E)/D_0$ (averaging also over the barrier regions, where no
channeled orbits exist, would reduce $ f^{\rm step}_{co}(B_0,V_0)$ by a factor of
2).  For $B_0=0$ we obtain
\begin{equation}
f_{co}^{\rm step }(0,V_0) = \frac{2}{\pi} \arcsin \sqrt{\frac{2 \, V_0
}{E_F + V_0}}  \, , \label{f_so_step}
\end{equation}
which for small $\epsilon = V_0 /E_F$ has the expansion
\begin{equation}
f_{co}^{\rm step } = \frac{2 \sqrt{2 \epsilon}}{\pi } - \frac{(\sqrt{2
    \epsilon} \, )
    ^{\, 3}}{ 6 \pi } + O (\epsilon ^{\, 5/2} ) \, . \label{f_so_step}
\end{equation}
 This algebraic dependence  on the modulation strength $V_0$ is not an
 artifact of the step model. For the cosine modulation $V(x)=-V_0 \cos qx$ one
 obtains, with $ f_{co}^{\rm cos}(B_0,V_0) =D_{co}^{\rm cos}(E_F) /D_0
 =\langle D_{co}^{\rm cos}(x,E_F) \rangle /D_0 $,
\begin{equation}
f_{co}^{\rm cos}(0,V_0)= \frac{2}{\pi} \int_0^a \frac{dx}{a} \arcsin
\sqrt{\frac{V_{0}- V(x) }{E_F - V(x)}}  \, , \label{f_so}
\end{equation}
with the expansion
\begin{equation}
f_{co}^{\rm cos } = \frac{4 \sqrt{2 \epsilon}}{\pi ^2} + \frac{(\sqrt{2
    \epsilon} \, )
    ^{\, 3}}{ 9 \pi ^2} + O (\epsilon ^{\, 5/2} ) \,  \label{f_so_cos}
\end{equation}
for small amplitudes.

For the step modulation, also the $B_0$-dependence of $f_{co}(B_0,V_0)$ can be
calculated analytically. We suppress the lengthy result for $0<B_0 <
B_{crit}^{\rm step}$ [cf. Eq.~(\ref{critstep})]
and give only the result for $B_0 > B_{crit}^{\rm step}$,
\begin{equation}
f_{co}^{\rm step } = \frac{4R}{a}\left[ \sqrt{2 \epsilon} -
\sqrt{1-\epsilon} \arcsin \sqrt{\frac{2\epsilon}{1+\epsilon}}\,
\right] \, . \label{f_so_b} 
\end{equation}

For the cosine model at $0<B_0 < B_{crit}^{\rm cos}$, we can calculate the
positions and 
the values of the local extrema of $x_0^-(x)$ [and similar for $x_0^+(x)$]
analytically, and also $D^{\rm cos}_{co}(x;E) $. Contributions to the average
of $D^{\rm cos}_{co}(x;E) $ come from the $x$-interval between the location 
$x_{max}$  of the local maximum and the value $x_{upper}$ satisfying
$x_0^-(x_{upper}) =x_0^-(x_{max})$. The value  $x_{upper}$ and the integral of
$D^{\rm cos}_{co}(x;E) $ between $x_{max}$ and  $x_{upper}$ are calculated
numerically. The results are plotted in Fig.~\ref{channeledfrac} for both the
cosine and the step model.
 Figure~\ref{channeledfrac}(a) shows the magnetic field dependence and (b)
shows the dependence on the modulation strength of $f_{co}(B_0,V_0)$, for
$V_0 \leq E_F$ in all cases.
For the cosine model at a  given value of the magnetic field $B_0$, channeled
orbits exist if $V_0 /E_F > (2/qR)[1-(qR)^{-2}]^{1/2}$. Their number increases
with increasing $V_0$. For  $V_0 \rightarrow E_F$,
$f_{co}(B_0,V_0)$ approaches, with vertical slope, a finite value, which
decreases with increasing $B_0$. No channeled orbits survive if $B_0$ becomes
so large that $2/qR \geq \sqrt{2}$. 
For the step model, the given $B_0$-value sets a critical modulation strength
$V_{crit}^{\rm step}$ given by $V_{crit}^{\rm step}/E_F
=(\pi/qR)[1-(\pi/2qR)^2]^{1/2}$. For $V_0 < V_{crit}^{\rm step}$ only skipping
orbits exist, whereas for $V_0 >  V_{crit}^{\rm step}$ also 
traversing orbits contribute to $f_{co}(B_0,V_0)$, which leads to a change of
the curvature of the curves in  Fig.~\ref{channeledfrac}(b) at  $V_0 =
V_{crit}^{\rm step}$. 
With increasing $V_0$, $f_{co}(B_0,V_0)$ increases and reaches at $V_0=E_F$,
also with vertical slope, the value
\begin{equation}
f_{co}^{\rm step}(B_0,E_F)=\frac{2}{\pi} \left[ \frac{\pi}{2} - \arcsin
\frac{b}{\sqrt{2}} + \frac{b/\sqrt{2}}{1+\sqrt{1-b^2/2}}\right] \, ,\label{fstepef}
\end{equation}
if $b \equiv \pi/qR < \sqrt{2}$, and $ f_{co}^{\rm
  step}(B_0,E_F)=(2/\pi)\sqrt{2}/b$ if $b> \sqrt{2}$.
 
With the normalization chosen in Fig.~\ref{channeledfrac}(a), one obtains for
the $B_0$ dependence of $f_{co}$
nearly the same curve for all values of the modulation
strength, $0< V_0 < E_F$. It can be shown analytically that, for weak
modulation and to lowest order in $V_0 /E_F$, a single curve should result for
each of the models. Numerically we find this to be true for $ V_0 /E_F <
0.2$. For the step model $f_{co}^{\rm step}(B_{crit}^ {\rm
  step}(V_0), V_0) / f_{co}^{\rm step} (0, V_0)$ approaches the values $2/3$
for $V_0 \rightarrow 0$ and $2/ \pi$ for  $V_0 \rightarrow E_F$.

The algebraic dependence of the number of channeled orbits on the modulation
strength leads to a similar algebraic $V_0$ dependence of their contribution
to the resistivity at $B_0=0$, as we demonstrate now.

\subsection{Resistance at $B_0=0$  }
We now evaluate, for $B_0=0$, the classical 
velocity-velocity-correlation formula for the conductivity in the extreme
nonlocal limit $\lambda q \rightarrow \infty$. In this limit we expect the
errors due to the violation of the continuity equation to be small, as we
discussed in section \ref{cleanlimit}.
In the limit $B_0=0$, the constant of motion is $\langle v_y \rangle$, the
average velocity in
$y$-direction. The ``drifting  orbits'' transform into orbits, which
are unbound in $x$-direction and lead to a periodic velocity ${\bf v} (x+a) =
{\bf v} (x)$, whereas the ``channeled orbits'' remain bound in 
$x$-direction within one period, and their velocity in $x$-direction, $v_x$,
vanishes on average. Then the conductivity tensor reduces to
\cite{Gerhardts96:11064}
\begin{equation}
\sigma_{\mu \nu} = \frac{\sigma _0}{v_F^2 /2} \int_0^a \frac{dx_0}{a}
\int_{-\pi}^{\pi} \frac{d \varphi}{2\pi} v_{\mu}(\varphi,x_0)
\bar{v}_{\nu}(\varphi,x_0) \, , \label{v_v_corr}
\end{equation}
where $\bar{v}_{\nu}(\varphi,x_0)$ is the average of a velocity component
along the trajectory starting (for an arbitrary value of $y_0$) at $(x_0,y_0)$
with velocity ${\bf v} =v_F(x_0) (\cos \varphi , \sin \varphi )$, where
$v_F(x_0) =v_F[1-V(x_0)/E_F]^{1/2}$. 

First, we consider a pure electric modulation, which implies that $v_y$ is
constant along any trajectory. Since the modulation potential $V(x)$ vanishes
on average, we see immediately that $\sigma _{yy} = \sigma _0$ and
$\sigma_{xy} =0$ have  the same Drude values as the unmodulated
system. According to  Eq.~(\ref{v_v_corr}), the channeled orbits
 do not contribute to $\sigma_{xx}$. For the trajectories extended in
 $x$-direction,  one has $\bar{v}_{x}(\varphi,x_0) =a / T(\varphi,x_0)$, where
\begin{equation}
T(\varphi,x_0)=\int_0^a \frac{dx}{\sqrt{v_F^2[1-V(x)/E_F]-v_y^2(
    \varphi,x_0)}} \label{time}
\end{equation}
is the time an electron needs to traverse one period of the modulation. Since $
T(\varphi,x_0)$ is an even function of $ \varphi$, Eq.~(\ref{v_v_corr}) yields
also $\sigma_{yx} =0$.  From Eqs.~(\ref{v_v_corr})
and (\ref{time}) we obtain for the electric cosine modulation
\begin{equation}
\frac{\sigma_{xx}^{\cos }}{\sigma_0} = (1+ \epsilon) \, \Phi \! \left(\frac{2
  \epsilon}{1+\epsilon} \right)   \, , \label{sig_cos}
\end{equation}
where $\epsilon =V_0 / E_F$,
\begin{equation}
\Phi( \eta )= \eta ^{3/2} \int_{\eta}^1 dt ~ [ \, t^2 \, {\rm K}(t) \,
\sqrt{t-\eta} \, ]^{-1}
\label{Phi}
\end{equation}
and ${\rm K}(t)$ is the complete elliptic integral of the first kind
\cite{Abramowitz}. Apparently $\sigma_{xx}^{\cos} \rightarrow 0$ for $V_0
\rightarrow E_F$. For $\epsilon \rightarrow 0$ one obtains a power expansion in
$\sqrt{\epsilon}$, the coefficients of which can be calculated numerically. To
leading order in $\epsilon$ one obtains
\begin{equation}
\frac{\Delta \rho_{xx}^{\cos } }{\rho_0} = 0.69454 \, \left(\frac{V_0}{E_F}
  \right)^{ 3/2} + O 
   ([V_0 /E_F ] ^{ \,  5/2} ) \,
  . \label{Drho_cos} 
\end{equation}

For the step model the corresponding results can be calculated
  analytically. The exact result for the conductivity is 
\begin{eqnarray}
 \frac{\sigma_{xx}^{\rm step }}{\sigma_0} =1 &-& \frac{3 - \epsilon}{\pi} \,
 \sqrt\frac{1 - \epsilon }{2 \epsilon} \label{sig_step} \\
 &+& \frac{3 - 2\epsilon - 5
 \epsilon ^2}{2 \pi \, \epsilon } \arcsin \sqrt{\frac{ 2 \epsilon}{ 1
 + \epsilon }}  \, , \nonumber
\end{eqnarray}
with the leading term of  the resistance correction 
\begin{equation}
\frac{\Delta \rho_{xx}^{step } }{\rho_0} = \frac{16}{15 \pi} \left(\frac{2
  V_0}{E_F}   \right)^{ 3/2} + O 
   ([V_0 /E_F ] ^{ \,  5/2} ) \,
  . \label{Drho_step} 
\end{equation}
Obviously, the step modulation yields the same half-integer power dependence of
the resistivity on the modulation strength. Moreover, if we choose the mean
values of the squared modulation potentials equal, i.e., replace $V_0$ in the
step potential by  $V_0 / \sqrt{2} $,
the right hand side of Eq.~(\ref{Drho_step}) becomes $0.571 \,
(V_0/E_F)^{3/2}$, where the prefactor is comparable with that in
Eq.~(\ref{Drho_cos}). 

The $B_0=0$ conductivities for pure magnetic modulations can be calculated
similarly. For the modulation $B_m(x) =B_m^0 \cos qx$ one obtains
\begin{equation}
\frac{\sigma_{xx}^{\cos }}{\sigma_0} = (1+\alpha)^2 \, \Phi \! \left(\frac{4
  \alpha}{(1+\alpha)^2} \right) \, , \label{sigmagnetcos}
\end{equation}
where $\Phi(\eta)$ is given by Eq.~(\ref{Phi}), and $\alpha = 1/q R_m
=\omega_m /qv_F$. For $\alpha \rightarrow 0$ this yields a power expansion in
$\sqrt{\alpha}$ similar to Eq.~(\ref{Drho_cos}),
\begin{equation}
 \frac{\Delta \rho_{xx}^{\cos } }{\rho_0} = 1.9645 \, (qR_m)^{-3/2} +
 O([qR_m]^{-2}) \, , \label{Dr_cos_mag}
\end{equation}
but now the correction term is larger. In this collisionless limit, the
conductivity vanishes if all trajectories become bound in $x$-direction. For
the cosine 
modulation this happens at $1/qR_m =1$. For the magnetic step modulation,
$B_m(x)= B_m^0 {\rm sign}(\cos qx)$ the corresponding condition is $a/4
R_m=1$, which means that a strip of constant $B_m(x)$ just can accommodate a
cyclotron orbit with diameter $2R_m$. As a function of $a/4R_m$, the
conductivity for the magnetic step modulation shows a similar behavior as that
for the magnetic cosine modulation as a function of $1/qR_m$.

It can be shown analytically that, for $B_0=0$, $\sigma_{yy} =\sigma_0
$ is not changed by the modulation. It is, however, interesting to calculate
the fraction $\Delta \sigma_{yy}^{co}/ \sigma_0$ contributed 
by the channeled orbits to $\sigma_{yy}$. This is easily shown to equal the
fraction contributed to the density,
\begin{equation}
\frac{\Delta \sigma_{yy}^{co}}{ \sigma_0} = \int_0^a \frac{dx}{a} \, \frac{
  n_{co}(x;E_F)}{\bar{n}_{el}} \, . \label{Dsigso} 
\end{equation} 
%


%
\newpage
\begin{figure}
\caption{(a) $S^{(0,0)}_q$ vs.~$1/qR \propto B_0$ for different values of the
  mean free path $\lambda $, given in terms of the modulation period $a = 2
  \pi /q$; (b) and (c):
  selected magnetoresistance results   for pure electric modulation
  $V(x)=V_0 \cos qx$ to order $V_0^2$,  calculated from
  Eq.~(\protect\ref{G_xx}). In (b) $\Delta \rho_{xx}$ is plotted
  in units of $\rho _1 = (\rho _0 /2) (V_0 /E_F)^2$, and the dash-dotted line
  represents the local limit for    $\lambda q $ = 1.0. In (c)  $\Delta
  \rho_{xx}$ is plotted 
  in units of $\rho _2 = \rho _0  (V_0 /E_F)^2 (\lambda q /2 )^2 $, and the
  thin line represents the approximation of
  Ref.~\protect\cite{Gerhardts96:11064}. 
}\label{beenakker}
\end{figure}

\begin{figure}
\caption{Magnetoresistivity vs.~$1/qR \propto B_0$ for (a) mixed electric and
  magnetic, and (b) mixed electric and mobility modulations, calculated from 
 Eqs.~(\protect\ref{G_xx})-(\protect\ref{Smunu}) for cosine modulations
 $V(x)=V_0 \cos qx$, $\omega_m (x)= \omega^0 _m \cos qx$, and $r(x)= r^0 _m
 \cos qx$, of comparable effective amplitudes $\alpha _V \equiv \lambda q
 V_0 /2   E_F$, $\alpha  _B \equiv \tau \omega^0 _m$ and  $\alpha  _{\mu}
 \equiv  \tau r^0 _m$,  respectively, with $\lambda q =30$.  $\gamma$ is a
 small, but otherwise arbitrary,  dimensionless constant.
}\label{beenakker2}
\end{figure}

\begin{figure}
\caption{Typical orbits (lower panel) in the electric superlattice potential
 $V(x)= 0.4 \, E_F \cos qx $, with $q = 2\pi /a$, for three values of the
 magnetic field $B_0$ corresponding to $\beta =1/qR = 0.25$, 0.1, and
 0.001. The corresponding values of the constant of motion $x_0$ are indicated
 in the upper panels by horizontal bars ending on the curves $x_0 ^{\pm} (x)$
 defining the turning points. Channeled orbits exist for $\beta > \beta
 _{crit} \approx 0.2 $.  }\label{orbits}
\end{figure}

\begin{figure}
\caption{Magnetoresistivity versus $1/qR \propto B_0$ for a weak electric
  cosine modulation with $V_0/E_F$=0.02 and $ \lambda q$=27.7. Solid line:
  numerical result, dotted line: analytical approximation
  (\protect\ref{G_xx}), dash-dotted: contribution of channeled orbits (see
  text). Inset: the same for magnetic cosine modulations with $1/qR_m $ =
  0.34, 0.15, and 0.05 (from top to bottom at arrow, for further explanations
  see text) and  $ \lambda q$=27.7. 
 }\label{fig5_11}
\end{figure}
\begin{figure}
\caption{Magnetoresistivity vs.~$1/qR \propto B_0$ for anisotropic scattering,
  modelled by Eq.~(\protect\ref{Pforward}), and an electric cosine modulation
  with $V_0/E_F =0.02$; $\tau_0 = 27.7/q v_F$. 
 } \label{fig5_35}
\end{figure}
\begin{figure}
\caption{Magnetoresistivity vs.~$1/qR$ for cosine modulation of the electric
  potential and the mobility, with $\bar{\tau} v_F q=27.7$. Solid lines: $\rho
  _{xx} 
  /\rho _0$ for pure electric modulation with $V_0/E_F =0.2$ and  $\rho _{yy}
 /\rho _0$ for pure mobility modulation with $r_m=1/ \bar{\tau}$; broken lines:
  mixed modulation.
 } \label{fig5_45}
\end{figure}
\begin{figure}
\caption{Fraction of channeled orbits, as function of magnetic field $B$ (a)
  and 
  modulation strength $V_0$ (b), for an electric cosine modulation $V(x)=-V_0
  \cos qx$ and a step modulation $V(x)=-V_0 \, {\rm sign} (\cos qx)$; (a):
  $B$-dependence for fixed $V_0 /E_F$ = 0.01 (solid lines) and 0.95
  (dash-dotted lines), (b): $V_0$-dependence for fixed $1/qR$ = 0.0, 0.05,
  0.2, and 0.4 (from top to bottom), with solid lines for the cosine and
  dash-dotted lines for the step modulation.
}     \label{channeledfrac}
\end{figure}

\end{document}